\documentclass[preprint,3p,12pt]{elsarticle}
\usepackage{mathrsfs}
\usepackage{amsmath}
\usepackage{stmaryrd}
\usepackage{bbding}
\usepackage{dcolumn}
\usepackage{graphicx}
\usepackage{amsfonts}
\usepackage{amssymb}
\usepackage{psfrag}
\usepackage{wrapfig}
\usepackage{subfigure}
\usepackage{makeidx}
\usepackage{bm}
\usepackage{epsf}
\usepackage{color}
\usepackage{epsfig}
\usepackage{setspace}
\usepackage{graphicx}
\usepackage{epstopdf}
\usepackage{psfrag}
\usepackage{subfigure}
\usepackage{multirow}
\usepackage{diagbox}
\usepackage{verbatim}

\usepackage{makecell}
\usepackage{float}
\usepackage{esint}
\usepackage{comment}
\usepackage{movie15}
\usepackage{hyperref}
\usepackage[ruled,linesnumbered]{algorithm2e}
\usepackage{algorithmicx}
\usepackage[noend]{algpseudocode}

\newcommand{\mathsym}[1]{{}}
\newcommand{\unicode}[1]{{}}
\begin{document}

\title{Unified gas-kinetic wave-particle method for multi-scale phonon transport}

	\author[HKUST1]{Hongyu Liu}
    \ead{hliudv@connect.ust.hk}
	
        \author[HKUST1]{Xiaojian Yang}
	\ead{xyangbm@connect.ust.hk}

    \author[add1]{Chuang Zhang}
\ead{zhangc520@hdu.edu.cn}

    	\author[XJTU]{Xing Ji}
	\ead{jixing@xjtu.edu.cn}
	
	\author[HKUST1,HKUST2,HKUST3]{Kun Xu\corref{cor1}}
	\ead{makxu@ust.hk}
	
	\address[HKUST1]{Department of Mathematics, Hong Kong University of Science and Technology, Clear Water Bay, Kowloon, Hong Kong}

    	\address[HKUST2]{Department of Mechanical and Aerospace Engineering, Hong Kong University of Science and Technology, Clear Water Bay, Kowloon, Hong Kong}

        	\address[HKUST3]{Shenzhen Research Institute, Hong Kong University of Science and Technology, Shenzhen, China}
	\cortext[cor1]{Corresponding author}

	\address[XJTU]{Shaanxi Key Laboratory of Environment and Control for Flight Vehicle, Xi'an Jiaotong University, Xi'an, China}

    \address[add1]{Institute of Energy, School of Sciences, Hangzhou Dianzi University, Hangzhou 310018, China}

\begin{abstract}

Over the past $7$ decades, the classical Monte Carlo method has played a huge role in the fields of rarefied gas flow and micro/nano scale heat transfer, but it also has shortcomings: the time step and cell size are limited by the relaxation time and mean free path, making it difficult to efficiently simulate multi-scale heat and mass transfer problems from the ballistic to diffusion limit.
To overcome this drawback, a unified gas-kinetic wave-particle (UGKWP) method is developed for solving the phonon Boltzmann transport equation (BTE) in all regimes covering both ballistic and diffusive limits.
This method is built upon the space-time coupled evolution model of the phonon BTE, which provides the framework for constructing a multi-scale flux at the cell interfaces.
At the same time, in order to capture non-equilibrium transport efficiently, the multi-scale flux comprises two distinct components: a deterministic part  for capturing the near-equilibrium or diffusive transport and a statistical particle part for recovering non-equilibrium or ballistic transport phenomena.
The UGKWP method exhibits remarkable multi-scale adaptability and versatility, seamlessly bridging the gap between the diffusive and ballistic transport phenomena.
In the diffusive limit, the present method naturally converges to the Fourier's law, with the diminishing particle contribution, whereas in the ballistic limit, the non-equilibrium flux is fully described by the free-streaming particles.
This inherent adaptability not only allows for precise capturing of both equilibrium and non-equilibrium heat transfer processes but also guarantees that the model adheres strictly to the underlying physical laws in each phonon transport regime.
A series of numerical tests fully demonstrate the excellent performance of the UGKWP method in all Knudsen regimes, where the time step and cell size are not constrained by the relaxation time and mean free path in the diffusive regime.
The present method is an efficient and accurate computational tool for simulating multiscale non-equilibrium heat transfer, and offering significant advantages over traditional methods in terms of numerical performance and physical applicability.

\end{abstract}


\maketitle

\section{Introduction}

Effective thermal management in micro- and nano-scale electronic devices is crucial due to the challenges posed by high levels of integration and power densities \cite{TCAD_application_intel_2021_review,murthy2005review,review_2019_CMOS,HUA2023chapter}.
At these length scales, the typical Fourier's law of heat conduction fails to capture non-equilibrium thermal conduction phenomena such as ballistic transport, nonlocal, nonlinear, size effects and complex interfacial scattering mechanisms~\cite{phononsnanoscale,RevModPhys.90.041002,RevModPhys.94.025002}.
Therefore, it is essential to develop heat conduction models that accurately reflect the underlying phonon dynamics, including drift, scattering, absorption, and emission to predict thermal behavior reliably. In this regard, the phonon Boltzmann transport equation (BTE) \cite{murthy2005review, ziman2001electrons, chattopadhyay2014comparative} emerges as a powerful tool, providing a comprehensive framework capable of addressing the limitations of Fourier's law and capturing the complex transport mechanisms dominant in modern electronic devices~\cite{TCAD_application_intel_2021_review,murthy2005review,review_2019_CMOS,HUA2023chapter}.

Unfortunately obtaining analytical solutions of the BTE for realistic electronic device configurations is highly challenging because of the multiscale properties of phonon transport and its complex interactions within materials.
Consequently, numerical simulations become indispensable, necessitating the development of advanced computational methods to accurately and efficiently model phonon transport for improved thermal management in electronic devices.
The BTE has seven dimensions: time, three spatial position dimensions, one frequency dimension, and two angular dimensions \cite{barry2022boltzmann, mazumder_boltzmann_2022}.
This high dimensionality results in enormous computational cost, posing significant challenges for numerical simulations~\cite{guo_progress_DUGKS, peraud_monte_2014, lacroix2005monte}.

Many numerical methods have been developed to solve the phonon BTE.
These can be broadly classified into two categories: deterministic methods and stochastic statistical methods.
Deterministic approaches include methods such as the lattice Boltzmann method \cite{GuoZl13LB}, explicit discrete ordinate method (DOM) \cite{stamnes1988dom, murthy1998domunstruct, SyedAA14LargeScale, FivelandVA96Acceleration}, (discrete) unified gas kinetic scheme (UGKS/DUGKS) \cite{guo_progress_DUGKS, guo2016dugksphonon, luo2017dugksphonon, zhang2019dugksphonon}, synthetic iterative scheme \cite{Chuang17gray, zhang2023acceleration, zhang2025synthetic}, among others.
On the other hand, stochastic statistical methods are exemplified by the Monte Carlo \cite{DSMC_book_1994,DSMC_phonon_1994, mazumder2001monte, lacroix2005monte,mittal2010monte, PJP11MC, peraud2015derivationmonte, peraud_monte_2014} method.
The LBM method is based on a near-equilibrium assumption \cite{qian1992lattice}, which makes it suitable for thermal conduction problems at low Knudsen numbers; however, its computations are not accurate for high Knudsen number regimes \cite{GuoZl13LB, chattopadhyay2014comparative}.
The DOM method discretizes the angular space numerically to capture non-equilibrium effects. However, this approach requires an enormous amount of memory in high-dimensional cases.
Moreover, the explicit DOM method decouples the scattering/collision and streaming processes of the distribution function, which necessitates that the time step should be smaller than the relaxation time.
Consequently, as the Knudsen number decreases, the computational cost of the DOM method becomes prohibitively high \cite{stamnes1988dom}.
To overcome the above shortcomings, the phonon transport and scattering are coupled together in a single time step in the DUGKS, so that it inherently captures the heat transfer physics across various regimes and its time step is not limited by the relaxation time \cite{guo_progress_DUGKS}.
However, this method still suffers from a significant memory overhead due to the high-dimensional discretization of the angular space.
The MC method uses statistical particles to replace the integration over a high-dimensional discretized angular space, significantly reducing memory usage and making it more efficient than the explicit DOM method in high Knudsen number regimes. However, since the MC method also decouples collisions from streaming just like the explicit DOM, the computational time step or grid size must be smaller than the relaxation time or mean free path due to the numerical stability and accuracy. As the Knudsen number gradually decreases, the computational cost of the MC method becomes prohibitively high due to frequent collisions\cite{peraud_monte_2014, lacroix2005monte, PJP11MC, SILVA2024108954, PATHAK2021108003}.
In addition, it suffers serious statistics errors.

To sum up, the deterministic methods suffer from enormous memory consumption, especially in the ballistic regime. The Monte Carlo methods demand a huge computational effort in the near-diffusive limit. The macroscopic diffusive equations cannot capture ballistic phonon transport.
Drawing on the advantages and disadvantages of the above methods, the particle version of UGKS \cite{UGKS}, known as unified gas-kinetic wave-particle (UGKWP), was developed, combining the macroscopic equations' efficiency in the diffusive regime with the effective non-equilibrium representation provided by Monte Carlo methods.
The UGKWP method has been widely applied to multiscale neutral gas transport \cite{zhu2019ugkwp}, plasmas \cite{liu2021ugkwp, pu2025ugkwp}, multiphase flows \cite{yang2024solid}, radiation \cite{yang2025rad} and turbulence simulation \cite{yang2025turb}.

Please note that the wave-particle approach discussed here differs from that in quantum mechanics~\cite{phononsnanoscale,kaviany_2008}. The phonon BTE describes only the particle aspect of phonon wave-particle duality, while the ``wave'' in this work refers to macroscopic temperature waves phenomenon.
In this paper, a regime-adaptive multiscale UGKWP method is developed based on the integral solution of the phonon BTE.
In this approach, the non-equilibrium transport process represented by the high-dimensional distribution function is replaced by the transport of statistical particles.
In the diffusive limit, the flux automatically reverts to typical Fourier's law, meaning that there are no particles in UGKWP; thus, both the computational cost and memory consumption are comparable to traditional methods for solving the Fourier's law.
Conversely, in the ballistic limit, the flux is entirely represented by the non-equilibrium transport of particles.
Compared with UGKS, UGKWP offers the advantage of regime adaptive in the angular space. Specifically, statistical particles are sampled at high Knudsen numbers to represent free transport. This significantly reduces memory overhead and improves computational efficiency compared to the high-dimensional velocity space used in UGKS.

This paper is organized as follows. Section 2 introduces the governing equations of the phonon transport system, and details the multi-scale UGKWP method for solving the phonon transport. Section 3 shows the numerical
results. The last section is the conclusion.

\section{Methodology}

This section will firstly introduce the phonon BTE as the governing equation. Then, we will detail the construction of the UGKWP method for phonon transport.

\subsection{Phonon Boltzmann Transport Equation}

The phonon BTE is an effective tool for describing the heat conduction in solid materials from the ballistic to diffusive regimes. In general, the equation can be simplified using the BGK type relaxation time approximation model \cite{BGK, ziman2001electrons, Chuang17gray, PJP11MC}
\begin{equation}
\frac{\partial f}{\partial t}+\boldsymbol{V_g} \cdot \nabla f=\frac{1}{\tau}\left(f^{e q}-f\right),
\label{eq:BGKBTE}
\end{equation}
where $f$ is the phonon distribution function, $\boldsymbol{V_g}=|\boldsymbol{V_g}| \boldsymbol{s}$ is the group velocity, $\boldsymbol{s}= (\cos \theta, \\ \sin \theta \cos \varphi, \sin \theta \sin \varphi)$ is the unit directional vector $(\theta$ is the polar angle and $\varphi$ is the azimuthal angle), $\tau$ is the relaxation time, $f^{eq}$ is the equilibrium distribution function, which satisfy the Bose-Einstein distribution,
\begin{equation}\label{bose-einstein}
f^{eq} (T)=\frac{1}{\exp \left(\hbar \omega / k_B T\right)-1},
\end{equation}
where $\hbar$ is the reduced Planck constant, $\omega$ is the frequency, $k_B$ is the Boltzmann constant and $T$ is the temperature.

In this paper, we employ the gray model, which neglects phonon dispersions and polarization characteristics and introduces the assumption that the phonon group velocity and relaxation time are constants.
Although this simplification cannot accurately capture frequency-dependent or anisotropic phonon transport behaviors in solid materials, it provides insights into phonon transport phenomena, such as ballistic transport and boundary scattering effects.
The Knudsen number of the system is defined as $\mathrm{Kn}=\lambda / L_0$, where $L_0$ is the characteristic length of the system and $\lambda=\boldsymbol{|V_g|} \tau$ is the phonon mean free path.
Eq.~\eqref{eq:BGKBTE} can also be expressed in terms of the phonon energy density per unit solid angle
\begin{equation}
\frac{\partial e}{\partial t}+\boldsymbol{V_g} \cdot \nabla e=\frac{1}{\tau}\left(e^{e q}-e\right).
\end{equation}
where
\begin{equation} \label{derivation function}
e(\boldsymbol{x}, \boldsymbol{s}, \omega, p)=
\sum_p \int   \left( \hbar \omega\left( f-f^{e q}\left(T_{\mathrm{ref}}\right)\right) D(\omega, p) / 4 \pi \right)   \mathrm{~d} \omega.
\end{equation}
\begin{equation} \label{derivation equilibrium}
e^{e q}(\boldsymbol{x}, \boldsymbol{s}, \omega, p)=\sum_p \int   \left( \hbar \omega\left(  f^{eq} (T) -f^{e q}\left(T_{\mathrm{ref}}\right)  \right) D(\omega, p) / 4 \pi \right)   \mathrm{~d} \omega.
\end{equation}
where $D(\omega, p)$ is the phonon density of state and $T_{\mathrm{ref}}$ is the reference temperature.
Taking a first-order Taylor expansion of distribution function at $T_0$, Eq.~\eqref{derivation equilibrium} can be expressed as
\begin{equation}\label{derivation final equilibrium}
    e^{eq} \approx \frac{C  \left(T-T_{\mathrm{ref}}\right)}{4 \pi},
\end{equation}
where $C $ is the volumetric specific heat.
The local energy $E$, temperature $T$ and heat flux $\boldsymbol{q}$ are obtained by taking the moments of the phonon distribution function of the energy density over the whole solid angle space,
\begin{align}
E  &=  \iint_{4 \pi} e \mathrm{~d} \Omega  ,  \\
T  &= \frac{ E }{C}  +T_{\mathrm{ref}} ,  \\
\boldsymbol{q} &= \iint_{4 \pi} \boldsymbol{V_g} e \mathrm{~d} \Omega.
\end{align}

\subsection{Multi-Scale UGKWP method}

The wave-particle approach discussed in this paper differs from that in quantum mechanics~\cite{phononsnanoscale}.
The phonon BTE describes only the particle aspect of phonon wave-particle duality, while the "wave" in this work refers to macroscopic temperature waves.
In the UGKWP method, the cell-averaged values are evolved using a finite volume approach. By performing an angular integration on phonon BTE and then applying finite volume spatial discretization, the energy evolution equation can be obtained:
\begin{equation}
     E_i^{n+1}-E_i^{n}+\frac{1}{V_i}\sum_{j \in N(i)} S_{i j} \boldsymbol{F}_{i j}=0,
\end{equation}
where $V_i$ is the volume of cell $i$,  $S_{ij}$ is the area of the $j$-th interface of cell $i$. $E_i$ is the associated cell-averaged energy. $F_{i j}$ denotes the macroscopic flux across the interface $S_{i j}$, which can be written as
\begin{equation}
    \boldsymbol{F}_{i j}=\int_0^{\Delta t}\iint_{4\pi} \boldsymbol{V_g} \cdot \boldsymbol{n}_{ij} e_{ij} \mathrm{~d} \Omega \mathrm{~d} t,
\end{equation}
where $\boldsymbol{n}_{ij}$ is the unit normal vector pointing out of the cell.

The key in constructing the interface multiscale flux lies in obtaining the distribution function at the interface; at the interface, the integral solution of the phonon BGK equation is:
\begin{equation} \label{integral gray}
e(\mathbf{0}, t, \boldsymbol{\Omega})=\frac{1}{\tau} \int_0^t e^{e q}\left(x^{\prime}, t^{\prime}, \boldsymbol{s}\right) e^{-\frac{t-t^{\prime}}{\tau}} \mathrm{d} t^{\prime}+e^{-\frac{t}{\tau}} e_0(-\boldsymbol{V_g} t, \boldsymbol{V_g}).
\end{equation}
This integral solution describes the transport process of particles arriving at the terminal point along characteristic lines. Specifically, particles with their initial distribution have a probability $e^{-\frac{t}{\tau}}$ of preserving that distribution while moving toward the terminal point. If a collision occurs at an intermediate time $t^{\prime}$, the particles have a probability $\frac{1}{\tau}e^{-\frac{t-t^{\prime}}{\tau}}$ of moving to the interface according to the local equilibrium distribution.
This process fundamentally differs from the DOM method. The interface distribution function, obtained through the integral solution, couples the particle migration and collision processes. Consequently, the method is not constrained by the mean free path, and the constructed interface flux is multiscale.
Furthermore, performing a first-order Taylor expansion in time and space along the characteristic line for $e^{e q}(\boldsymbol{x}, t)$ as Eq.~\eqref{taylor-equlibrium}:

\begin{equation} \label{taylor-equlibrium}
    e^{e q}(\boldsymbol{x}, t)=e_0^{e q}+e_t^{e q} t+e_x^{e q} \cdot \boldsymbol{V_g}t.
\end{equation}
Substituting it into Eq.~\eqref{integral gray}, the explicit formula for distribution function at time $t$ can be obtained as Eq.~\eqref{time-dependent-distribution-function}
\begin{equation} \label{time-dependent-distribution-function}
e(t)=c_1 e_0^{e q}+c_2 e_x^{e q} \cdot \boldsymbol{V_g}+c_3 e_t^{e q}+c_4 e_0+c_5 e_{\boldsymbol{x}} \cdot \boldsymbol{V_g}.
\end{equation}
Taylor expansion in space for the initial distribution function is also considered in Eq.~\eqref {time-dependent-distribution-function}
with the coefficients
$$
\begin{aligned}
& c_1=1-e^{-t / \tau} \\
& c_2=t e^{-t / \tau}-\tau\left(1-e^{-t / \tau}\right) \\
& c_3=t-\tau\left(1-e^{-t / \tau}\right) \\
& c_4=e^{-t / \tau} \\
& c_5=-t e^{-t / \tau}.
\end{aligned}
$$
Then, the multi-scale macro numerical flux passing the interface can be written as:
\begin{equation}
\begin{aligned}
\int_0^{\Delta t} \iint_{4\pi} \boldsymbol{V_g} \cdot \boldsymbol{n}_{i j} e_{i j}^{\prime \prime}(t)  \mathrm{~d} \Omega \mathrm{~d} t= & \iint_{4\pi} \boldsymbol{V_g} \cdot \boldsymbol{n}_{i j}\left(q_1 e_0^{e q}+q_2 e_{\boldsymbol{x}}^{e q} \cdot \boldsymbol{V_g}+q_3 e_t^{e q}\right) \\
& +\boldsymbol{V_g} \cdot \boldsymbol{n}_{i j}\left(q_4 e_0+q_5 e_{\boldsymbol{x}} \cdot \boldsymbol{V_g}\right) \mathrm{~d} \Omega  \\
= & \mathcal{F}_{i j}^{e q}+\mathcal{F}_{i j}^{f r},
\end{aligned}
\end{equation}
with the coefficients
$$
\begin{aligned}
& q_1=\Delta t-\tau\left(1-e^{-\Delta t / \tau}\right), \\
& q_2=2 \tau^2\left(1-e^{-\Delta t / \tau}\right)-\tau \Delta t-\tau \Delta t e^{-\Delta t / \tau}, \\
& q_3=\frac{\Delta t^2}{2}-\tau \Delta t+\tau^2\left(1-e^{-\Delta t / \tau}\right), \\
& q_4=\tau\left(1-e^{-\Delta t / \tau}\right), \\
& q_5=\tau \Delta t e^{-\Delta t / \tau}-\tau^2\left(1-e^{-\Delta t / \tau}\right) .
\end{aligned}
$$

Due to the symmetry of the integration domain in angular space, the equilibrium flux is reduced to the form given by Eq.~\eqref{wave-q2}

\begin{equation}\label{wave-q2}
\mathcal{F}_{i j}^{e q}=\iint_{4\pi} \boldsymbol{V_g} \cdot \boldsymbol{n}_{i j}\left(q_2 e_{\boldsymbol{x}}^{e q} \cdot \boldsymbol{V_g}\right) \mathrm{~d} \Omega.
\end{equation}

Moreover, $\boldsymbol{|V_g|}$ in phonon transport using the gray model is a constant. Thus, $\boldsymbol{V_g}=\boldsymbol{|V_g|} \boldsymbol{s},\boldsymbol{s}=(\cos \theta, \sin \theta \cos \varphi, \sin \theta \sin \varphi)$.
Meanwhile, we can get the micro derivative of $e^{eq}$ using the chain rule:
\begin{equation}\label{micro-derivative}
    \frac{\partial e^{eq}}{\partial x}=\frac{\partial e^{eq}}{\partial E} \frac{\partial E}{\partial x}=\frac{1}{4\pi}\frac{\partial E}{\partial x}.
\end{equation}
Thus, we have $e_{\boldsymbol{x}}^{e q}=\frac{1}{4 \pi} \nabla_x E$, then substituting it into Eq.~\eqref{wave-q2}, we can obtain:
\begin{equation}
\mathcal{F}_{i j}^{e q}=q_2 \frac{1}{4 \pi} \int_{\Omega} \boldsymbol{V_g} \cdot \boldsymbol{n}_{i j}\left(\nabla E \cdot \boldsymbol{V_g}\right) \mathrm{~d} \Omega.
\end{equation}
More specifically:
\begin{equation}\label{feq}
\begin{aligned}
\mathcal{F}_{i j}^{e q}&=\frac{q_2 \boldsymbol{|V_g|}^2}{4 \pi} \iint_{4\pi} \boldsymbol{s} \cdot \boldsymbol{n}_{i j}\left(\nabla E \cdot \boldsymbol{s}\right) \mathrm{~d} \Omega \\
&=\frac{q_2 \boldsymbol{|V_g|}^2}{3} \boldsymbol{n}_{i j} \cdot \nabla E.
\end{aligned}
\end{equation}
Recall the compatibility condition used in UGKS:
\begin{equation}\label{capability-condition}
    \iint_{4\pi} e^{eq} \mathrm{~d} \Omega = \iint_{\boldsymbol{V_n}>0} e^{eq,L} \mathrm{~d} \Omega +\iint_{\boldsymbol{V_n}<0} e^{eq,R} \mathrm{~d} \Omega.
\end{equation}
Then take spatial derivative of Eq.~\eqref{capability-condition} on both side:
\begin{equation}\label{capability-condition-derivative}
e_x^{e q}=\frac{1}{4 \pi}\left(\iint_{\boldsymbol{V_n}>0} e_{x}^{e q,L}  \mathrm{~d} \Omega+\iint_{\boldsymbol{V_n}<0} e_{x}^{e q,R}  \mathrm{~d} \Omega\right),
\end{equation}
which can be further simplified as:
\begin{equation}
e_x^{e q}=\frac{e_{x}^{e q,L}+e_{x}^{e q,R}}{2}.
\end{equation}
As a result, the macroscopic flux of the equilibrium part is:
\begin{equation}
\mathcal{F}_{i j}^{e q}=\frac{q_2 \boldsymbol{|V_g|}^2}{6}\left(\frac{\partial E^L}{\partial n}+\frac{\partial E^R}{\partial n}\right).
\end{equation}

For non-equilibrium transport, the free propagation of the initial state’s distribution function along the characteristic lines contributes to the non-equilibrium transport flux.
In the UGKS method, the distribution functions are stored in each cell. This allows the UGKS method to directly employ spatial reconstruction to obtain the initial distribution functions on both the left and right sides of the cell interface, thereby obtaining the non-equilibrium transport flux.
UGKP method employs statistical particles instead of the initial distribution function to characterize non-equilibrium transport efficiently. The free propagation of these particles along the characteristic lines represents the non-equilibrium transport flux. The particle trajectory is described as Eq.~\eqref{particle-path}

\begin{equation}\label{particle-path}
\boldsymbol{x}=\boldsymbol{x}^n+\boldsymbol{V_g} t_f,
\end{equation}
where $t_f$ denotes the phonon's free streaming time.
Recall that the integral solution indicates that a particle has a probability $e^{-\frac{t}{\tau}}$ of maintaining its initial state as it propagates from the initial time to time t.
Thus, the particle's free transport time can be sampled as follows:

\begin{equation}
t_f=\min \left[-\tau \ln (\eta), \Delta t\right] ,
\end{equation}
where $\eta$ represents a random number uniformly distributed over the interval $[0, 1]$.

Therefore, the evolution of the macroscopic energy can be described by the following equation:

\begin{equation}
E_i^{n+1}-E_i^n=-\frac{1}{V_i} \sum_{j \in N(i)} S_{i j} \mathcal{F}_{i j}^{e q} + \frac{1}{V_i}\left(\sum w_p^{in} - \sum w_p^{out}\right) .
\end{equation}

Based on the free transport time sampled for the particles, they can be divided into collisionless and collisional particles.
Collisionless particles refer to those that are transported without collisions during the physical time step $\Delta t$, meaning their free transport time is exactly $\Delta t$. In contrast, within the same physical time step $\Delta t$, collisional particles first undergo free transport for a duration $t_f$ and then experience a series of collisions.
Thus, the collisional particles are removed at the transport endpoint upon completing free transport, and their corresponding energy is counted into the relevant cell's total macroscopic energy.
From the updated cell total energy, the unsampled  particles are:
\begin{equation}
    E^h = E^{n+1} -\frac{1}{V_i} \sum w^{fr},
\end{equation}
where $E^h$ is the summation of the umsampled particles' energy, $w^{fr}$ is the energy of the collisionless particle.
Moreover, $E^h$ can be recovered by the particle re-sampling process as Eq.~\eqref{kp-sample} illustrates:

\begin{equation}\label{kp-sample}
    w^p=\frac{E^hV_i}{N},
\end{equation}
where N is the re-sample number, $w^p$ is the particle's energy.

The method described above is the UGKP method, in which, as previously stated, all the free-transport fluxes are entirely represented by statistical particles.
The UGKWP method is an extension of the UGKP method that more efficiently represents non-equilibrium effects. Unlike the UGKP method, which represents the entire non-equilibrium transport using particles, the UGKWP method decomposes the non-equilibrium transport into the wave and particle components, further reducing the number of sampled particles.

In UGKP, the resampled particles' free transport flux can be analytically written as:

\begin{equation}
\begin{aligned}
    F_{i j}^{f r, U G K S}\left(e^{eq,h}\right)&=\int_0^{\Delta t}\iint_{4\pi} \boldsymbol{V_g} \cdot \boldsymbol{n}_{ij} e^{-\frac{t}{\tau}} e^{eq,h}\left(-\boldsymbol{V_g}t,\boldsymbol{V_g}\right) \mathrm{~d} \Omega \mathrm{~d} t \\
    &= \iint_{4\pi} \boldsymbol{V_g} \cdot \boldsymbol{n}_{ij}  \left[ q_4 e^{eq,h}\left(0,\boldsymbol{V_g}\right) + q_5 \boldsymbol{V_g} \cdot e_x^{eq,h}\left(0,\boldsymbol{V_g}\right)\right]\mathrm{~d} \Omega.
\end{aligned}
\end{equation}

Meanwhile, the UGKWP method only resamples collisionless particles to maintain non-equilibrium transport.
Thus, the energy for collisionless particles is:
\begin{equation}
    e^{hp}=e^{-\frac{\Delta t}{\tau}}E^h.
\end{equation}
In a statistical sense, the free transport of these collisionless particles is equivalent to the flux contributed by the distribution function carrying the corresponding energy in the DOM (Discrete Ordinates Method) framework, as Eq.~\eqref{fr-dvm} shows:

\begin{equation}\label{fr-dvm}
\begin{aligned}
    F_{i j}^{f r, D O M }\left(e^{eq,hp}\right)&=e^{-\frac{\Delta t}{\tau}}\int_0^{\Delta t}\iint_{4\pi} \boldsymbol{V_g} \cdot \boldsymbol{n}_{ij}  e^{eq,h}\left(-\boldsymbol{V_g}t,\boldsymbol{V_g}\right) \mathrm{~d} \Omega \mathrm{~d} t \\
    &= e^{-\frac{\Delta t}{\tau}}\iint_{4\pi} \boldsymbol{V_g} \cdot \boldsymbol{n}_{ij}  \left[  \Delta t e^{eq,h}\left(0,\boldsymbol{V_g}\right) -\frac12 \Delta t^2 \boldsymbol{V_g} \cdot  e_x^{eq,h}\left(0,\boldsymbol{V_g}\right)\right]\mathrm{~d} \Omega.
\end{aligned}
\end{equation}

Therefore, the portion of the non-equilibrium flux arising from the wave component, excluding the free transport contribution of collisionless particles, is given by:

\begin{equation}
\begin{aligned}
F_{i j}^{f r, w a v e} & =F_{i j}^{f r, U G K S}\left(e^{eq,h}\right)-F_{i j}^{f r, D O M}\left(e^{eq,h p}\right) \\
& =\iint_{4\pi} \boldsymbol{V_g} \cdot \boldsymbol{n}_{i j}\left[\left(q_4-\Delta te^{-\frac{\Delta t}\tau}\right) e^{eq,h}(\mathbf{0}, \boldsymbol{V_g})+\left(q_5+\frac{\Delta t^2}{2} e^{-\frac{\Delta t}{\tau}}\right) \boldsymbol{V_g} \cdot e_x^{eq,h}(\mathbf{0}, \boldsymbol{V_g})\right] \mathrm{~d} \Omega.
\end{aligned}
\end{equation}

Moreover, $F_{i j}^{f r, w a v e}$ can be calculated explicitly as:

\begin{equation}\label{fr,wave}
    F_{i j}^{f r, w a v e}=\left(q_4-\Delta t e^{-\Delta t / \tau}\right) \frac{\boldsymbol{V_g}\left(E_L^h-E_R^h\right)}{4}
    + \left(q_5+\frac{\Delta t^2}{2} e^{-\Delta t / \tau}\right)\left[\frac{\boldsymbol{V_g}^2}{6}\left(\frac{\partial E^h}{\partial n}\right)^R+\frac{c^2}{6}\left(\frac{\partial E^h}{\partial n}\right)^L\right].
\end{equation}

Finally, the evolution of the UGKWP method for cell i can be written as:

\begin{equation}
E_i^{n+1}-E_i^n=-\frac{1}{V_i} \sum_{j \in N(i)} S_{i j} \left(\mathcal{F}_{i j}^{e q}+\mathcal{F}_{i j}^{fr,wave}\right)+\frac{1}{V_i}\left(\sum w_p^{i n}-\sum w_p^{\text {out }}\right).
\end{equation}

Further description of particle re-sampling and deletion processes is necessary to provide a more comprehensive introduction to the UGKWP method.
In this study, a reference number method \cite{zhu2019ugkwp} is employed for particle sampling to control the number of sampled particles.
Given the reference number, the reference sampling energy for a particle can be written as:

\begin{equation}
w_r=\frac{\left(E-E^h\right)+E^h e^{-\Delta t / \tau}}{N_r} V,
\end{equation}
where $w_r$ is the reference sampling energy for a particle, $N_r$ is the reference number.

As illustrated above, the exact re-sampling energy from the wave is $E^he^{-\frac{\Delta t}{\tau}}$. So, the exact sampling number is:

\begin{equation}
N_s= \begin{cases}0, & \text { if } V e^{-\frac{\Delta t}{\tau}} E^h \leq e_{\min } \\ 2\left\lceil\frac{e^{-\frac{\Delta t}{\tau}} E^h}{2\left(E-E^h\right)+2 E^h e^{-\Delta t / \tau}} N_r\right\rceil, & \text { if } V e^{-\frac{\Delta t}{\tau}} E^h>e_{\min }\end{cases},
\end{equation}
As a result, the energy of each particle is $\frac{e^{-\Delta t / \tau} E^h V}{N_s}$ and $e_{min}=0.0001E$.
Additionally, the velocity directions of these particles are uniformly distributed over the spherical domain.
This can be obtained by the Inverse Sampling method:

\begin{equation}
\theta=\arccos \left(1-2 \eta_1\right),
\end{equation}

\begin{equation}
\varphi=2 \pi \eta_2,
\end{equation}
where $\eta_1$ and $\eta_2$ are random numbers uniformly distributed in $[0,1]$.

Two methods are usually employed to treat isothermal boundary conditions on particles: the ghost cell method and the sampled particle flux on the face method.
For the ghost cell method, the energy $E^h$ within the ghost cell equals $E$, corresponding to the energy of the isothermal boundary. Particle sampling is carried out in the same manner as for the interior
cells; however, once a particle transports into the ghost cell, it is immediately removed.
Another method is more general and more convenient for the unstructured mesh. The total sampling energy for particles on faces can be analytically written as:
\begin{equation}\label{inflow-boundary}
\int_{\boldsymbol{V_n}<0} \boldsymbol{V_g} \cdot \boldsymbol{n}_{ij}  e^{e q}(T_b) \Delta t e^{\frac{-\Delta t}{\tau}} \mathrm{~d} \Omega.
\end{equation}
This means the free transport non-equilibrium flux from 0 to $\Delta t$. The inverse Sampling method is employed to sample the particle's velocity direction based on Eq ~.\eqref{inflow-boundary}.
Simultaneously, the energy flux is equally distributed among these sampled particles based on the equilibrium.

\begin{algorithm}\label{coarsening algorithm}
	\caption{UGKWP method for phonon transport}
	\label{cell-merge}
	Initialization: set initialize temperature and energy field\;
	  \textbf{while} ($t<t_{stop}$) \textbf{do}\;
        classify the remaining particles by compare $t_f$ and $\Delta t$\;
        Sample free stream particles both for the inner region and boundary\;
        spatial reconstruction for $E$ and $E^h$\;
        Calculate macroscopic flux $\mathcal{F}_{i j}^{e q}$ and $\mathcal{F}_{i j}^{f r, w a v e}$\;
        Particle free streaming\;
        Update Macroscopic energy by evolution equation\;
        Delete collisional particles which $t_f<\Delta t$\;
        \textbf{end while}\;
        Output
\end{algorithm}	

The asymptotic behavior and the computational cost of the UGKWP method for phonon transport are introduced.

In the collisionless limit (ballistic region) $\tau \rightarrow \infty$, which means the particle free-streaming time goes to infinity:
\begin{equation}
    t_f=\lim _{\tau \rightarrow \infty}(-\tau \ln (\eta)) \rightarrow \infty.
\end{equation}
As mentioned above, at the beginning of each time step, the remaining particles steaming time is $\min \left(\Delta t, t_f\right) = \Delta t$, which means collisionless in this new time step. Meanwhile, the newly sampled collisionless particle's free streaming time is also $\Delta t$.
So, at this limit, the UGKWP method solved the collisionless phonon BTE.

When $\tau \rightarrow 0$ (diffusive limit), which has $\Delta t \gg \tau$, yielding $e^{-\frac{\Delta t}{\tau}} \rightarrow 0$. for time coefficient $q_2$, $q_4$ and $q_5$, we have:
\begin{equation}
\begin{aligned}
        q_2 &=-\tau \Delta t + O(\tau^2) \\
        q_4 &=\tau\\
        q_5 &=\tau^2 + O(e^{-\frac{\Delta t}{\tau}}).
\end{aligned}
\end{equation}
Substituting it into the macroscopic flux as Eq.~\eqref{feq}
and Eq.~\eqref{fr,wave} and since $E^h=Ee^{-\frac{\Delta t}{\tau}}$ which is exponentially small, yielding $q_4$ and $q_5$ flux are exponentially small and no particle will be sampled,  we have:
\begin{equation}
\begin{aligned}
\mathcal{F}_{i j}^{an} & =\frac{-\tau \Delta t\left|\boldsymbol{V}_g\right|^2}{4 \pi} \iint_{4 \pi} \boldsymbol{s} \cdot \boldsymbol{n}_{i j}\left(\nabla E \cdot \boldsymbol{s}\right) \mathrm{d} \Omega +  O(\tau^2)\\
& =\frac{-\tau \Delta t\left|\boldsymbol{V}_{\boldsymbol{g}}\right|^2}{3} \boldsymbol{n}_{i j} \cdot \nabla E +  O(\tau^2) \\
&= \frac{-\tau \Delta tC_v\left|\boldsymbol{V}_{\boldsymbol{g}}\right|^2}{3} \boldsymbol{n}_{i j} \cdot \nabla T +  O(\tau^2) \\
&= \Delta t \kappa \boldsymbol{n}_{i j} \cdot \nabla T +  O(\tau^2).
\end{aligned}
\end{equation}
In this diffusive limit, the UGKWP method returns to Fourier heat conduction, and its computational costs are the same as those of the traditional method.

\section{Numerical Tests}
This section uses a series of benchmark heat transfer cases to validate the effectiveness of the proposed multiscale method for solving the phonon BTE.
The time step is determined by $\Delta t=\mathrm{CFL} \times \frac{\Delta x}{\boldsymbol{|V_g}|}$ with $\mathrm{CFL}=1.0$.
The results obtained by the present method will be
compared with the data predicted by the DUGKS \cite{zhang2019dugksphonon}.
The van Leer limiter is used in 1D and 2D cases to ensure numerical stability.
Moreover, $C_v$ and $\boldsymbol{|V_g|}$ is 1 in 1D and 2D cases  if not specified.

\subsection{Heat conduction across a ﬁlm}
In this section, the one-dimensional heat conduction in a dielectric film with a thickness of $L = 1$ is simulated, as illustrated in Fig.\ref{1d-heat}.

\begin{figure}[htb]	\label{1d-heat}
	\centering	
	\includegraphics[height=0.45\textwidth]{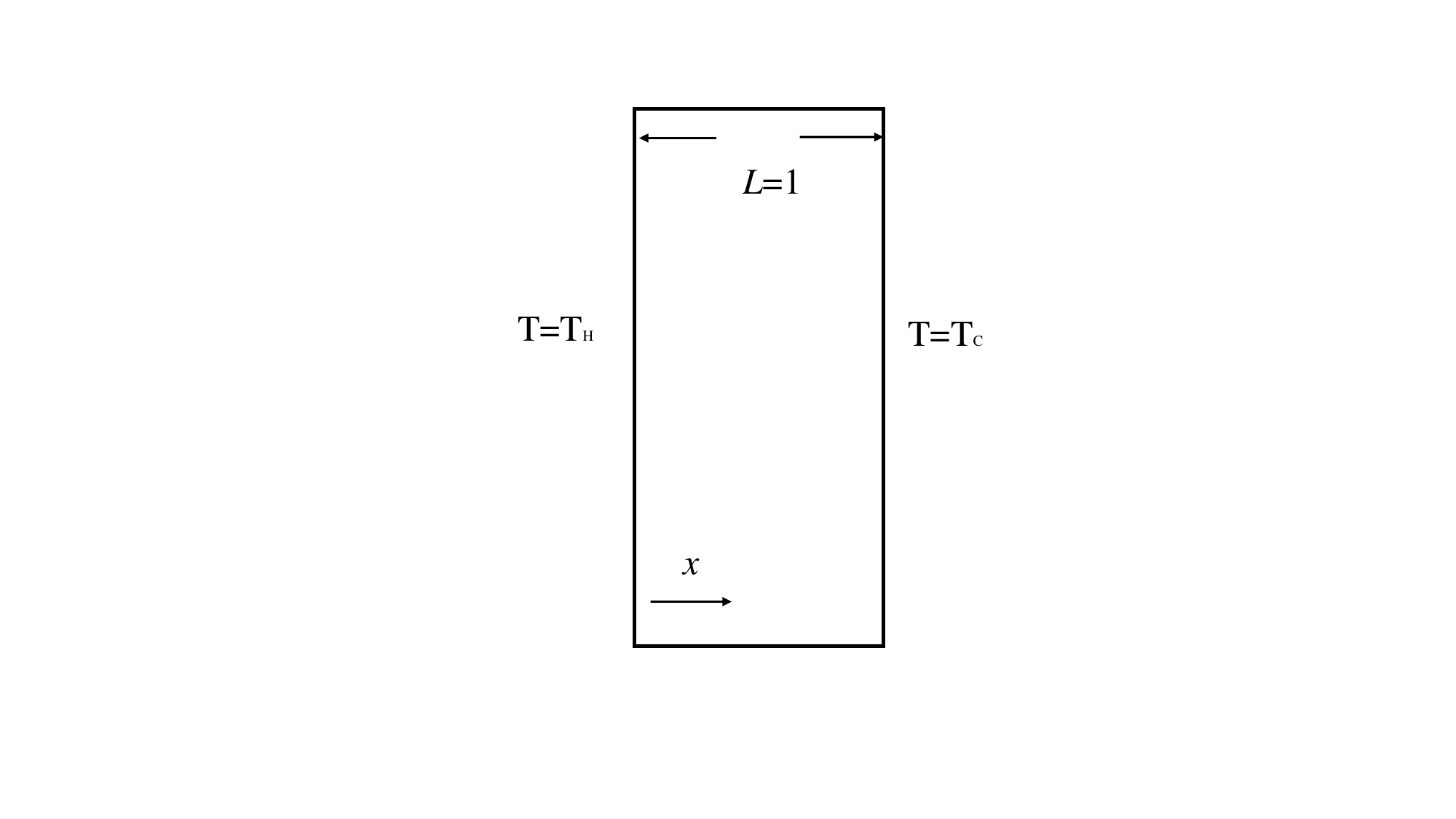}
	\caption{\label{1d-heat}
		A schematic diagram of one-dimensional heat conduction in a dielectric film.}
\end{figure}
At the left boundary $(x = 0)$, an isothermal high-temperature boundary is imposed with a temperature of $T_H$, while at the right boundary $(x = L)$, an isothermal low-temperature boundary is imposed with a temperature of $T_L$.
An analytical solution can be obtained in \cite{heaslet1965radiative-1d-ana-1, majumdar1993microscale-1d-ana-2}.

In this test case, the one-dimensional heat conduction problem is computed for Knudsen numbers of 10.0, 1.0, 0.1, and 0.01. The corresponding reference particle numbers in each cell are 3000, 800, 200, and 20 to balance computational efficiency with statistical noise.
The one-dimensional computational domain is discretized into 40 uniform cells, and the CFL number is set to 0.5.

Fig.\ref{1d-heat-result} illustrates a comparison among the computed results of the UGKWP method, the analytical solution, and those obtained by other methods under different Kn conditions, where $T^*=\frac{T-T_C}{T_H-T_C}$, $x^*=\frac{x-x_L}{x_R-x_L}$.
The results indicate that the UGKWP method agrees well with both the analytical solution and the DUGKS method. The only discrepancy is the statistical noise observed on the high-temperature side, which can be attributed to the fact that particles in this region carry higher energy, thereby amplifying the noise effect compared to the low-temperature side.
Moreover, when the Knudsen number is small, good agreement with the reference solution can be achieved using only 20 particles.
This observation further demonstrates that, in the continuum limit, the computational cost of the current method can be reduced to the same order of magnitude as that of the central scheme to solve Laplace's equation.

\begin{figure}[htb]	\label{1d-heat-result}
	\centering	
	\includegraphics[height=0.35\textwidth]{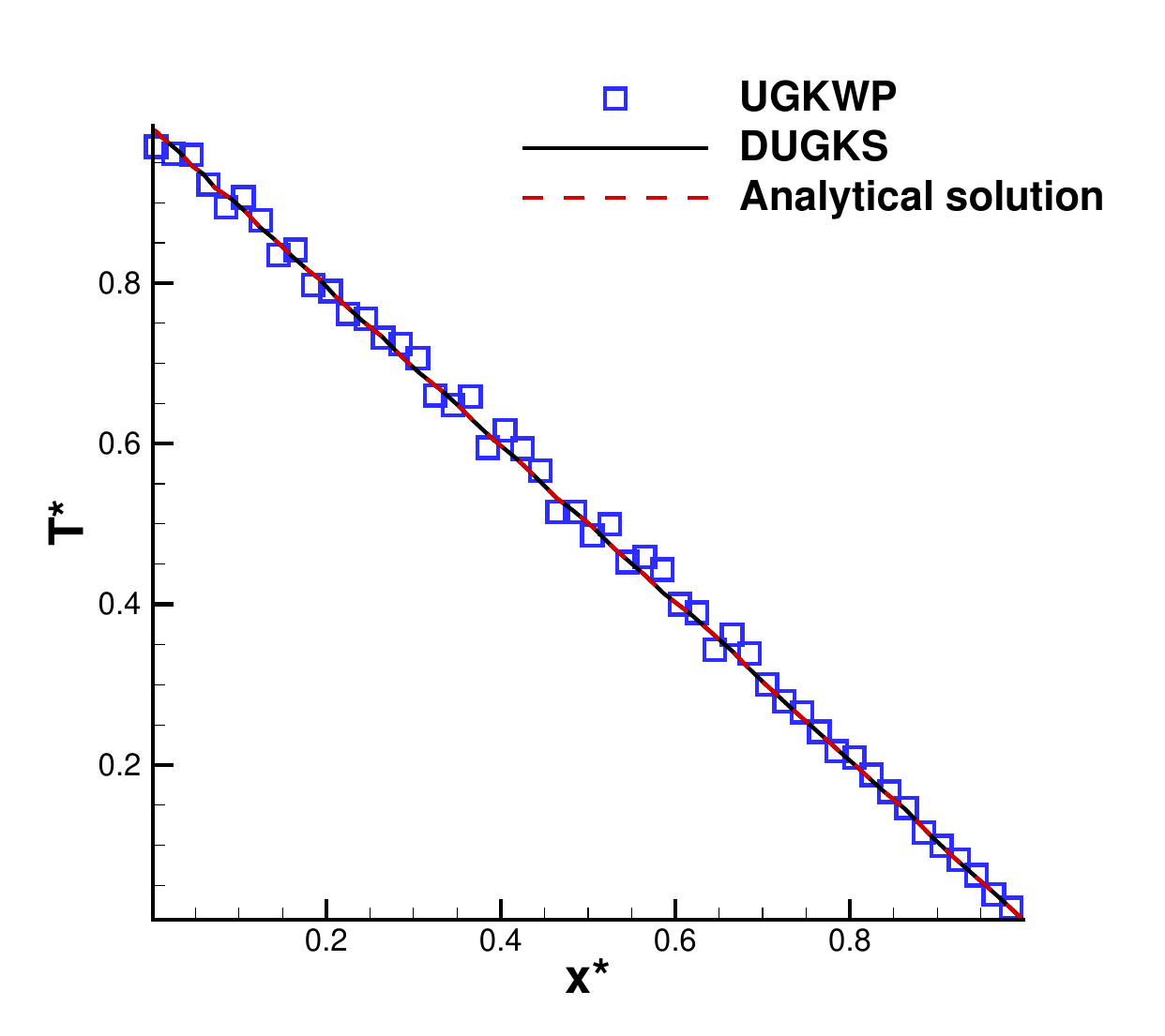}
        \includegraphics[height=0.35\textwidth]{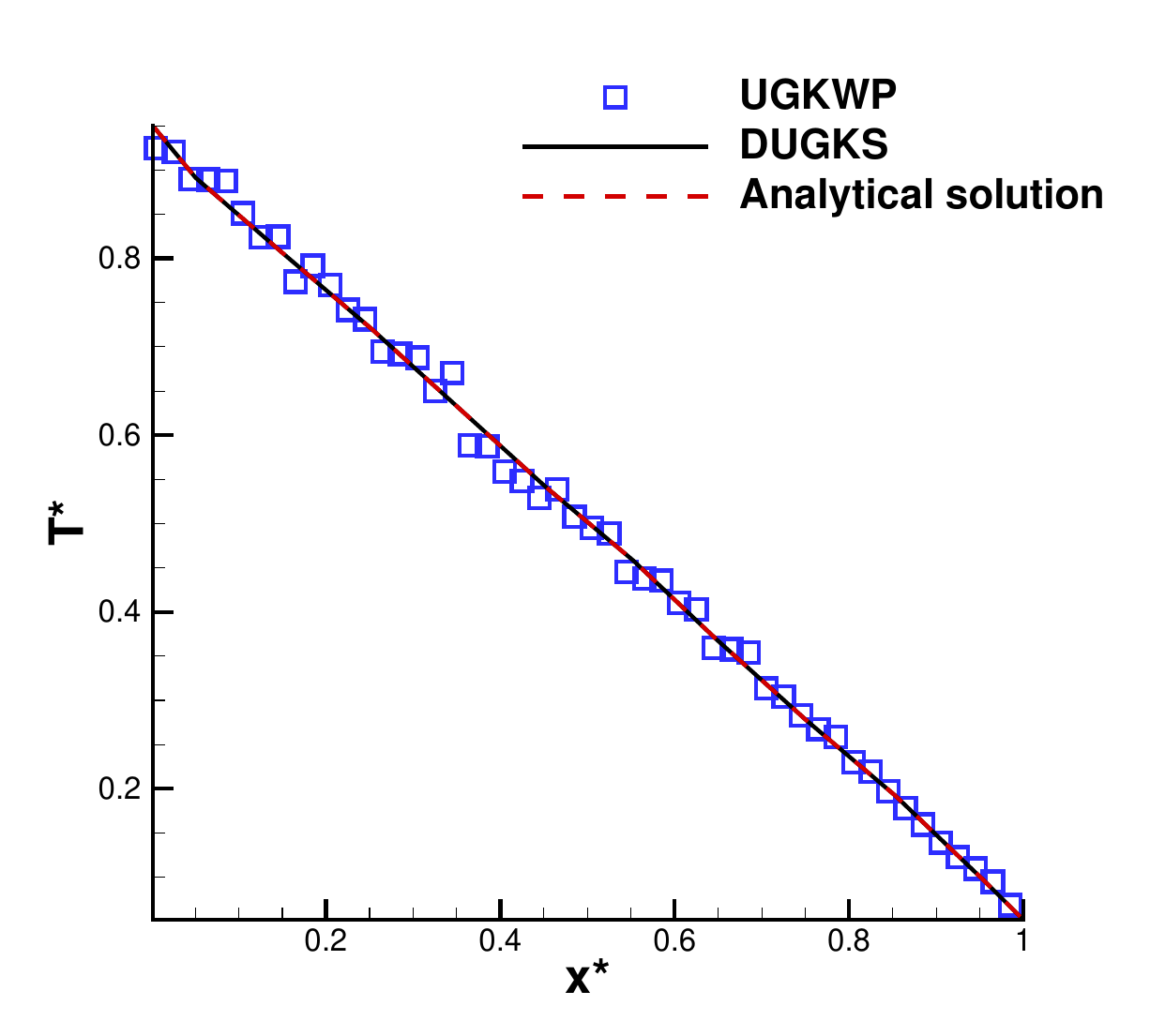}
        \includegraphics[height=0.35\textwidth]{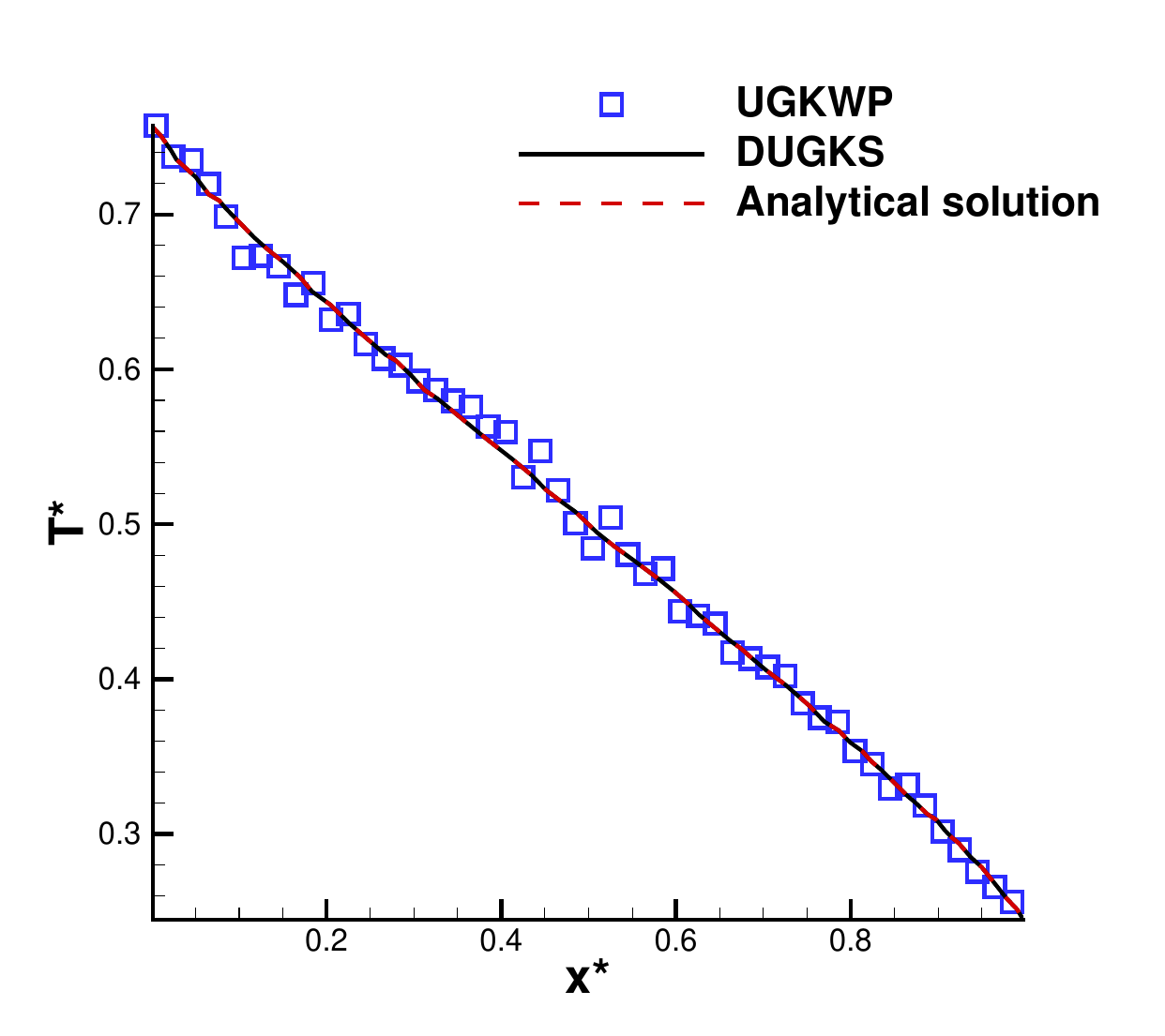}
        \includegraphics[height=0.35\textwidth]{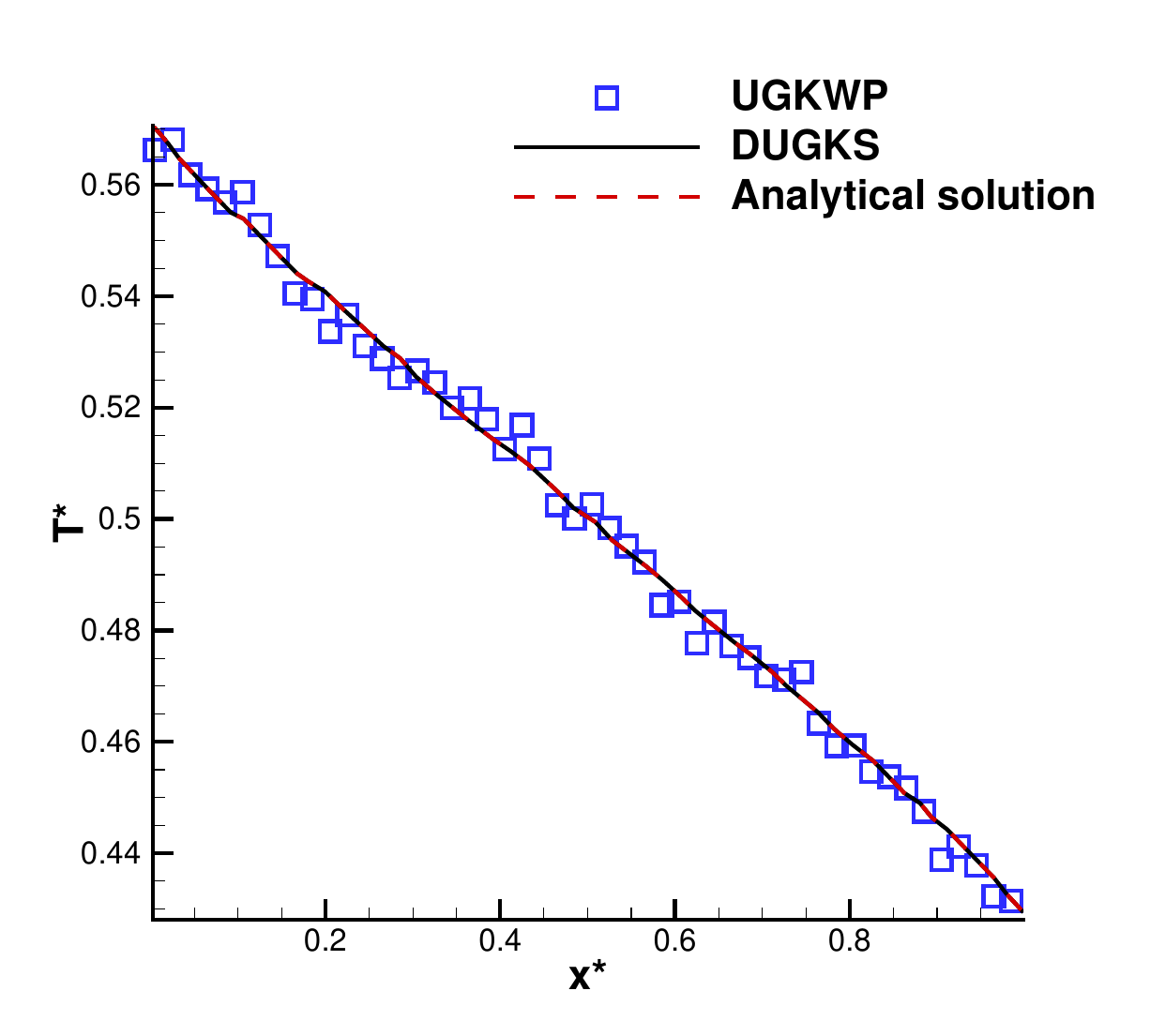}
	\caption{\label{1d-heat-result}
		Comparison of 1D heat conduction results across a ﬁlm, from Kn=0.01 to Kn=10.0. }
\end{figure}

\subsection{One-dimensional multiscale heat transfer problem}
To further demonstrate the capability of the proposed method in capturing multiscale non-equilibrium phenomena in phonon transport, a one-dimensional multiscale heat conduction case is designed in this section.
The computational setup and boundary conditions are identical to those in the first test case, except the relaxation time is now a spatially varying function, as detailed below.
\begin{equation}
\tau(x)=10^{A_1 \sin (2 \pi x / L)-A_2}.
\end{equation}
Here, $A_1$ and $A_2$ are adjustable constants. When $A_1$ is 1.5 and $A_2$ is -0.5, the value of $\tau$ spans the range $[10^{-1}, 10^{2}]$. This indicates that the Knudsen number varies by three orders of magnitude over the entire spatial domain, posing a significant challenge for multiscale methods. Consequently, this example is an excellent test case for evaluating the performance of the present method.
Initially, the temperature throughout the entire computational domain is uniformly set to $0.5·(T_H + T_c)$, representing the arithmetic mean of the high and low-temperature values. The reference sampling number for particles is 400, and the grid number is 400.
The result is shown in Fig.\ref{multiscale-1d-heat-result}, illustrating that the UGKWP method can automatically recover the heat transfer physics in different scales.

\begin{figure}[htb]	\label{multiscale-1d-heat-result}
	\centering	
	\includegraphics[height=0.40\textwidth]{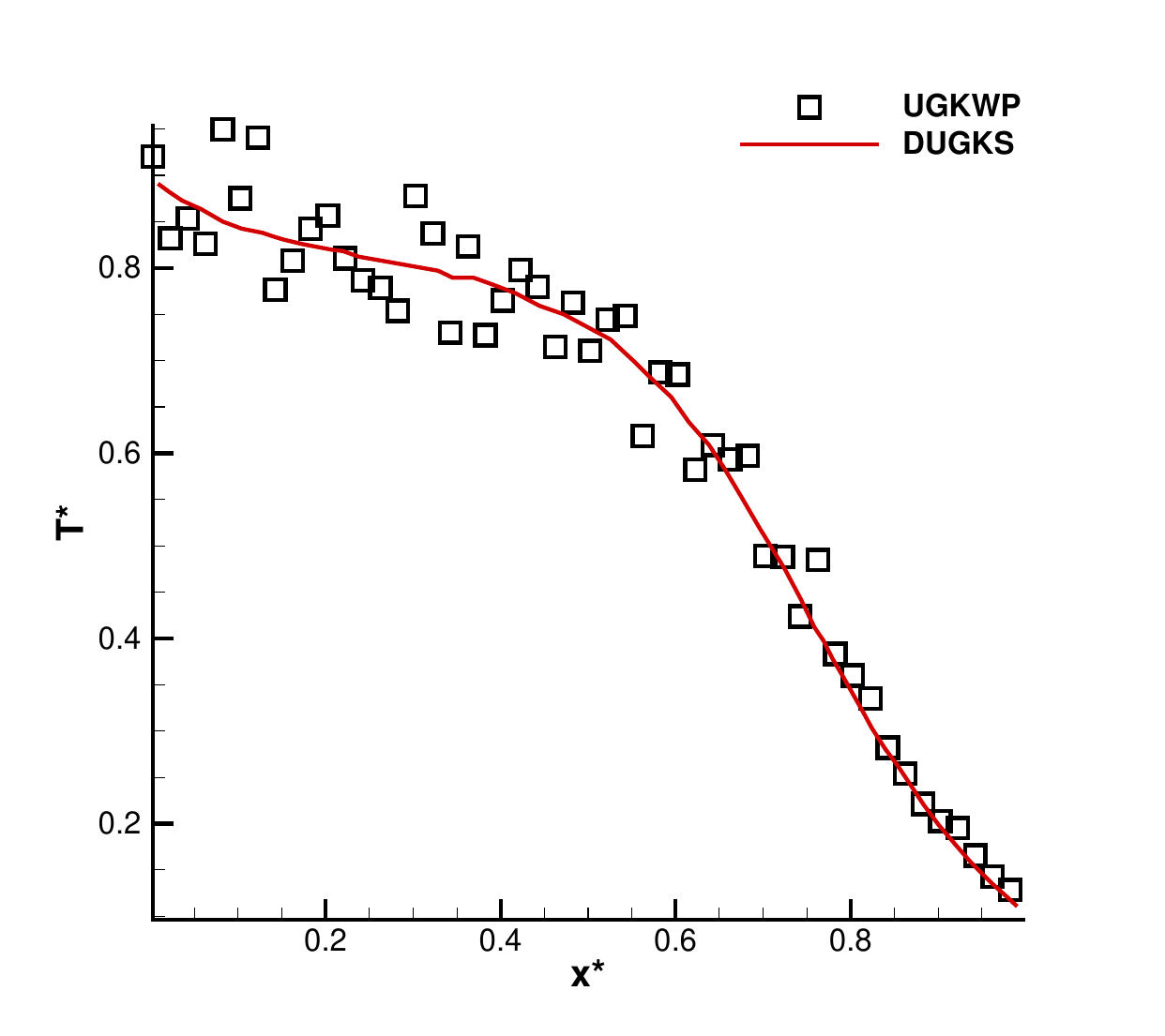}
        \includegraphics[height=0.40\textwidth]{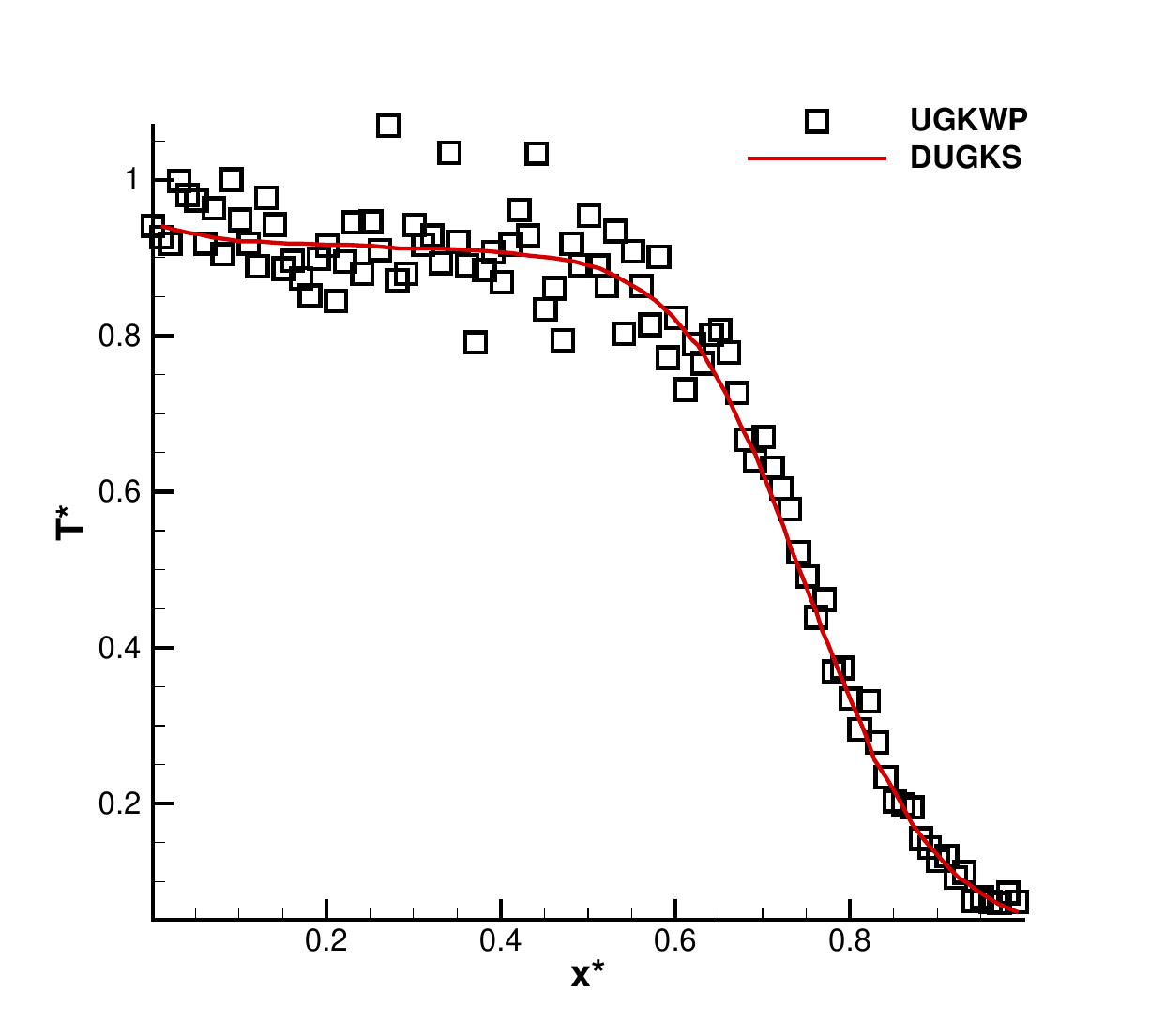}
        \includegraphics[height=0.40\textwidth]{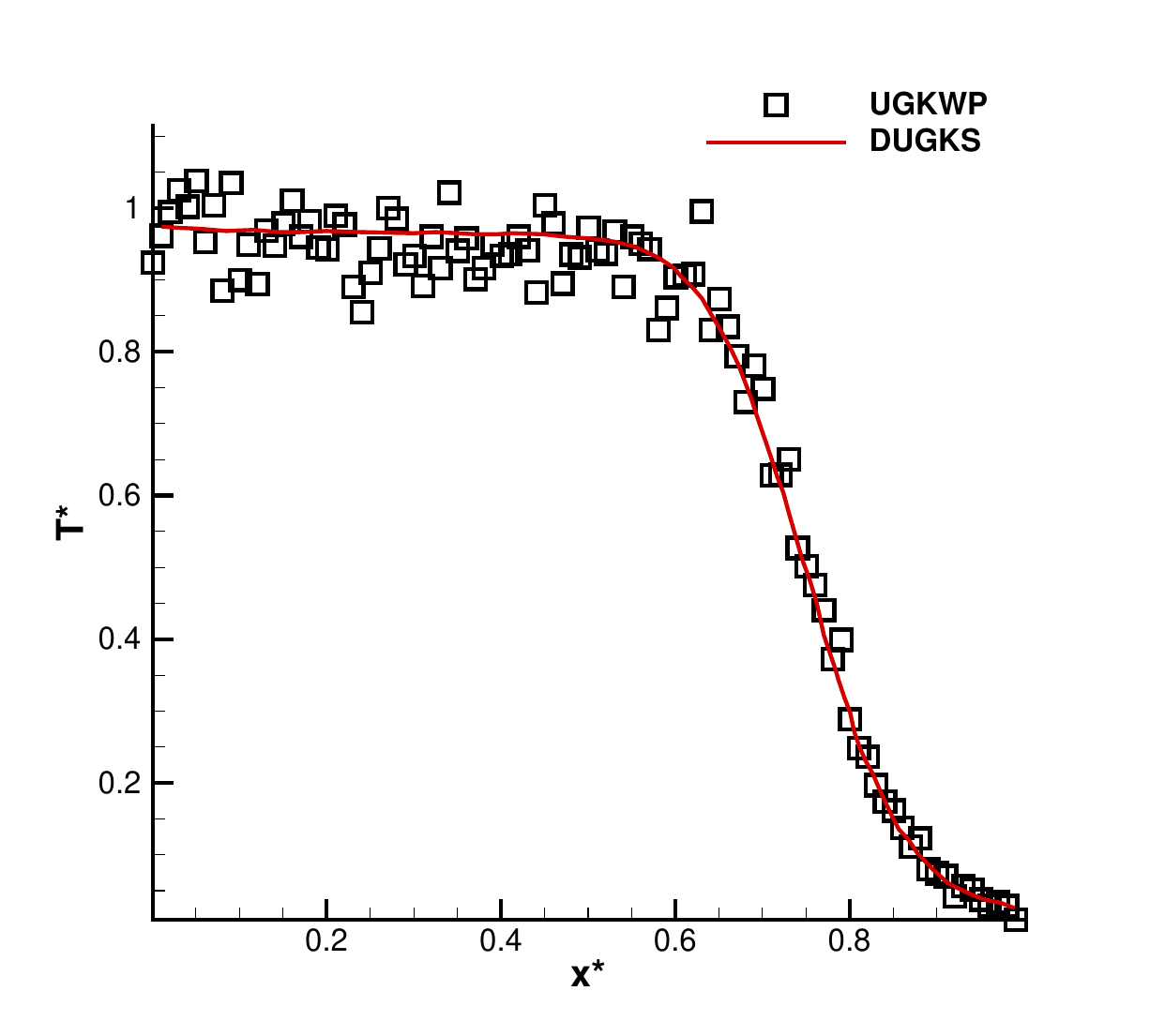}
	\caption{\label{multiscale-1d-heat-result}
		Comparison results of 1D multi-scale heat conduction across a ﬁlm, respectively corresponding: $A_1 = 0.5$, $A_1 = 1.0$, $A_1 = 1.5$. }
\end{figure}

\subsection{Transient thermal grating}
Transient thermal grating represents a sophisticated methodology for examining the thermal properties of materials and is frequently employed in laser heating experiments. The technique involves irradiating the material’s surface with a brief pulsed laser, thereby inducing localized temperature variations. More specifically, interference from crossed laser beams generates a spatially cosine-modulated temperature profile within the sample.
\begin{equation}
T(x, 0)=T_b+A_0 \cos (\alpha x),
\end{equation}
where $T_b$ is the background temperature, $A_0$ is the amplitude of the temperature variation, $\alpha = \frac{2\pi}{L}$ is
the wave number with $L = 1$ being the spatial grating period as an example to simulate the evolution process of peak temperature amplitude with time.
The periodic boundary conditions are applied at the boundary of $x = 0$ and $x = L$.

Numerical simulations are performed under different Knudsen numbers and compared with the analytical solutions derived in Ref .\cite{collins2013non-1d-unsteady},
\begin{equation}
A^*\left(t^*\right)=\operatorname{sinc}\left(\xi t^*\right) \exp \left(-t^*\right)+\int_0^{t^*} A^*\left(t^{\prime}\right) \operatorname{sinc}\left(\xi\left(t^{\prime}-t^*\right)\right) \exp \left(t^{\prime}-t^*\right) d t^{\prime},
\end{equation}
where $A^*=A / A_0, t^*=t / \tau, \xi=2 \pi \mathrm{Kn}$.
In this section, we simulated the cases including $\mathrm{Kn}=0.001$, $\xi = 0.25, 1.0,2.0$. This means a wide range of Kn numbers is covered, and the grid number is 100, and the reference sampling number for particles is 800 to reduce the statistical noise.  Results are shown in Fig.\ref{1d-transient}, illustrating that the UGKWP method can capture the phonon transport well from the diffusive regime to the ballistic regime.

\begin{figure}[htb]	\label{1d-transient}
	\centering	
	\includegraphics[height=0.40\textwidth]{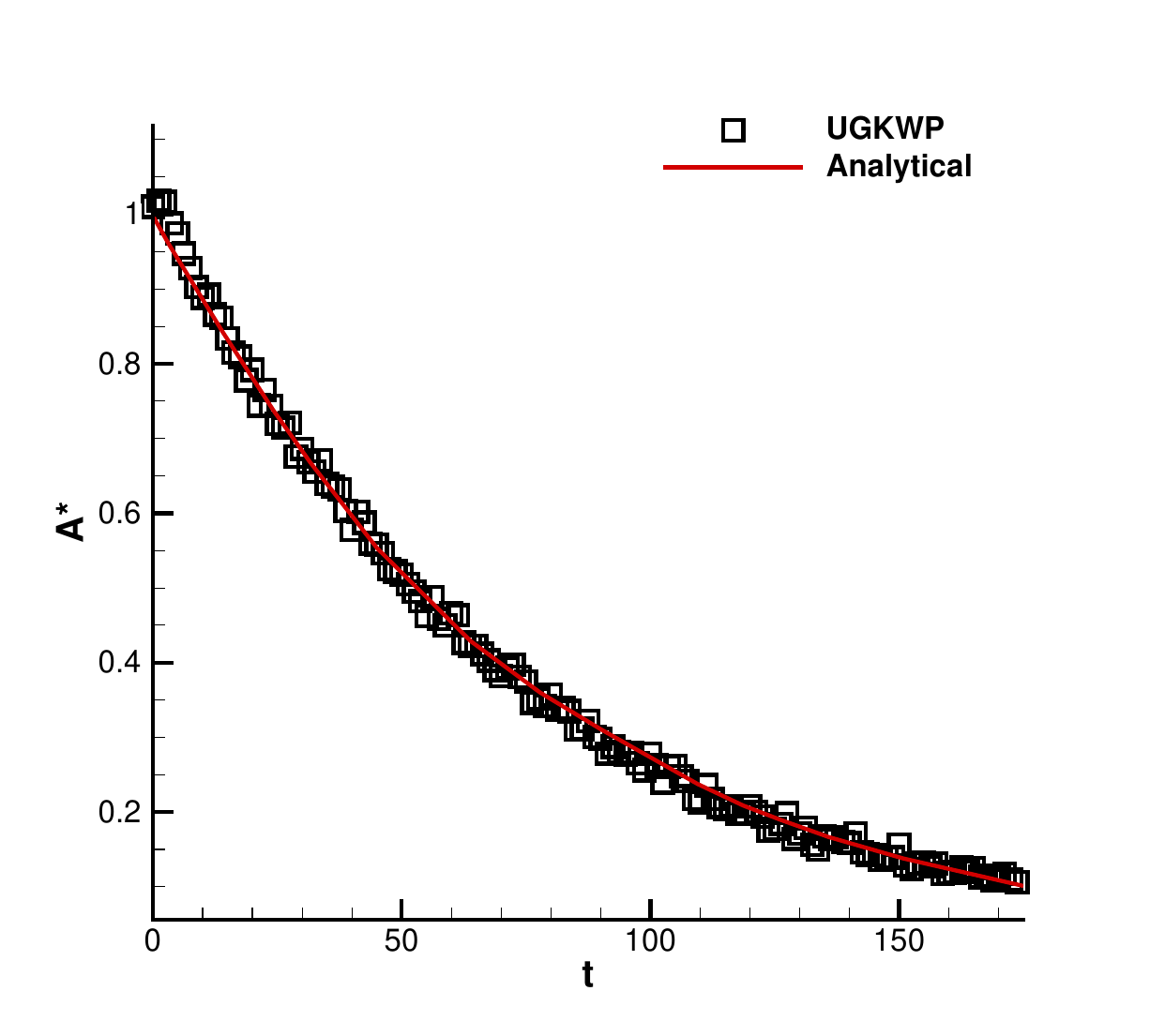}
        \includegraphics[height=0.40\textwidth]{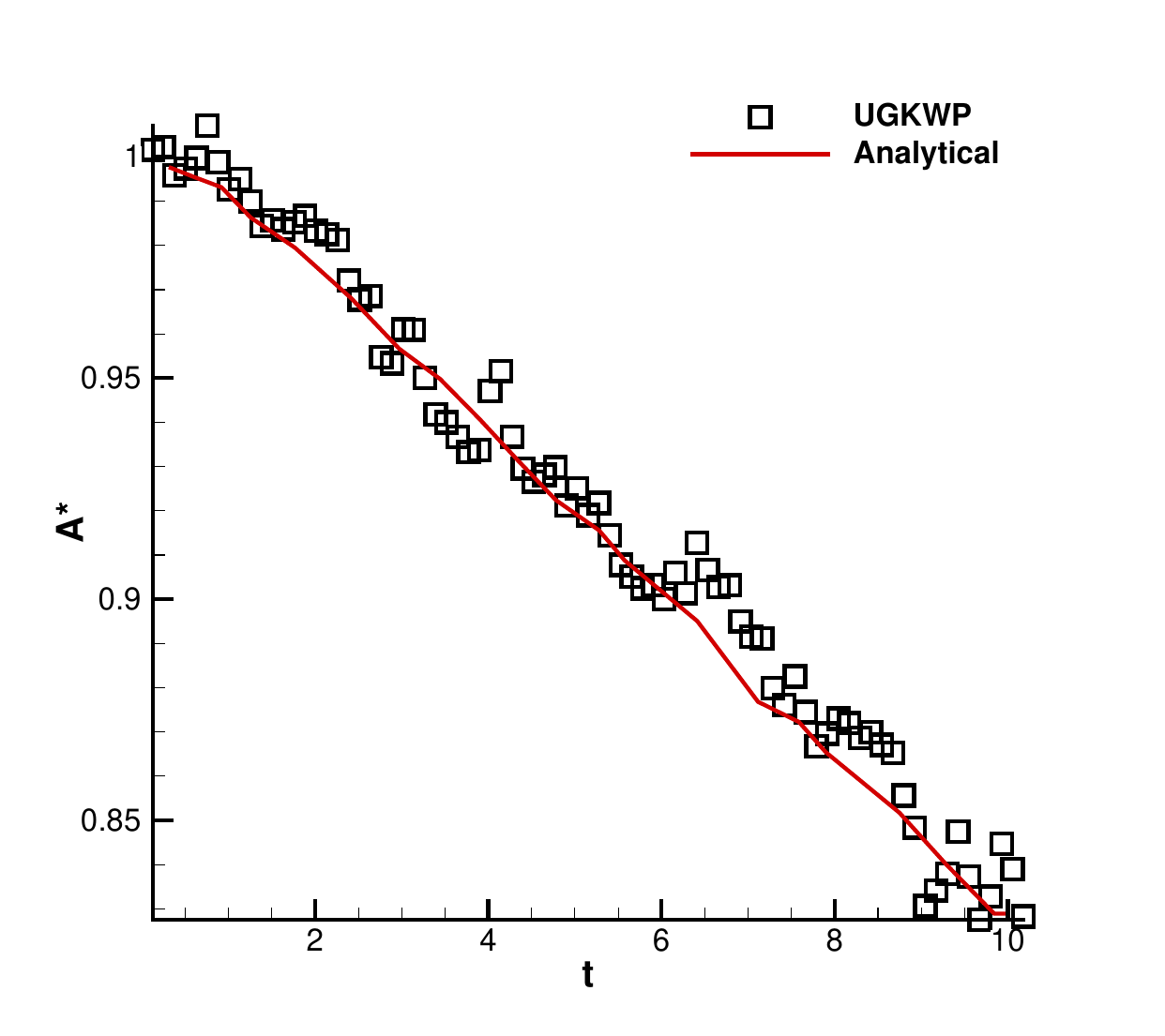}
        \includegraphics[height=0.40\textwidth]{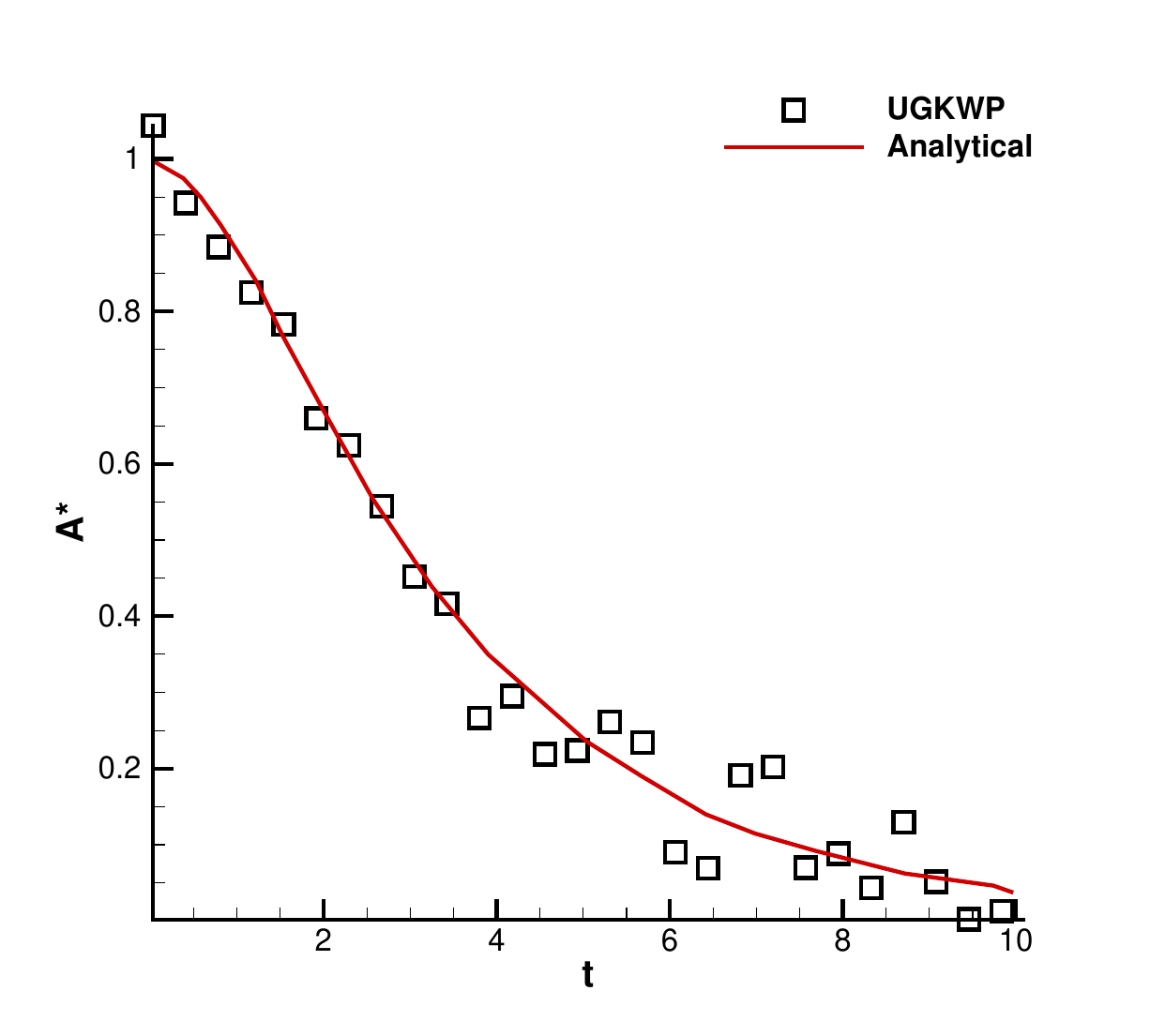}
        \includegraphics[height=0.40\textwidth]{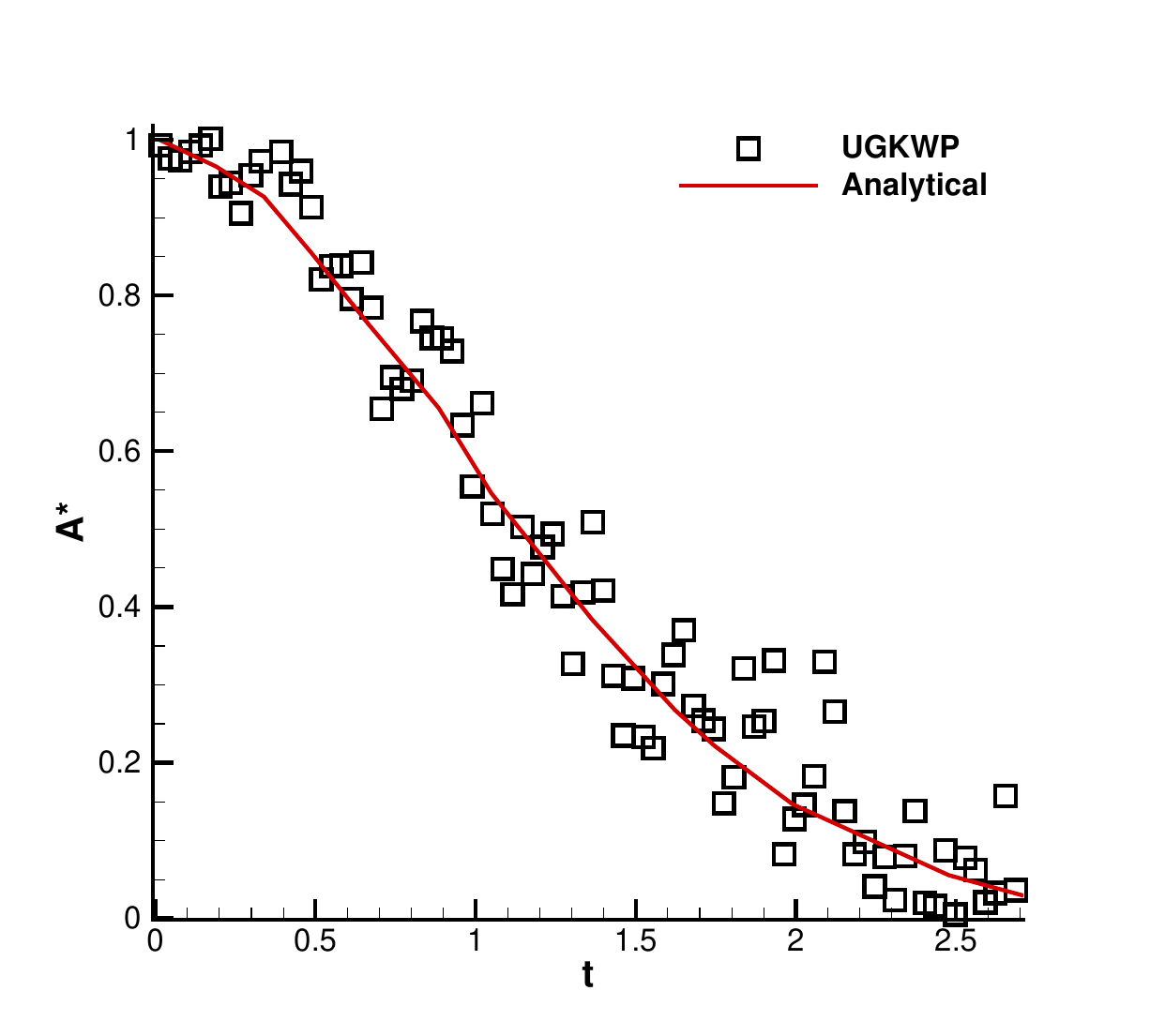}
	\caption{\label{1d-transient}
		Comparison results of 1D transient thermal grating, respectively corresponding: Kn = 0.001, $\xi = 0.25$, $\xi = 1.0$, $\xi=2.0$. }
\end{figure}

\subsection{Heat transfer in the 2D square domain}
To validate the effectiveness of the UGKWP method for phonon transport in multidimensional physical space, this section investigates the heat conduction problem in a 2D square domain across a range of Knudsen numbers.
Specifically, the top boundary is maintained at a high temperature $T_H$, while the remaining boundaries are held at a lower temperature $T_C$ as illustrated in Fig.\ref{2d-square}.

\begin{figure}[htb]	\label{2d-square}
	\centering	
	\includegraphics[height=0.45\textwidth]{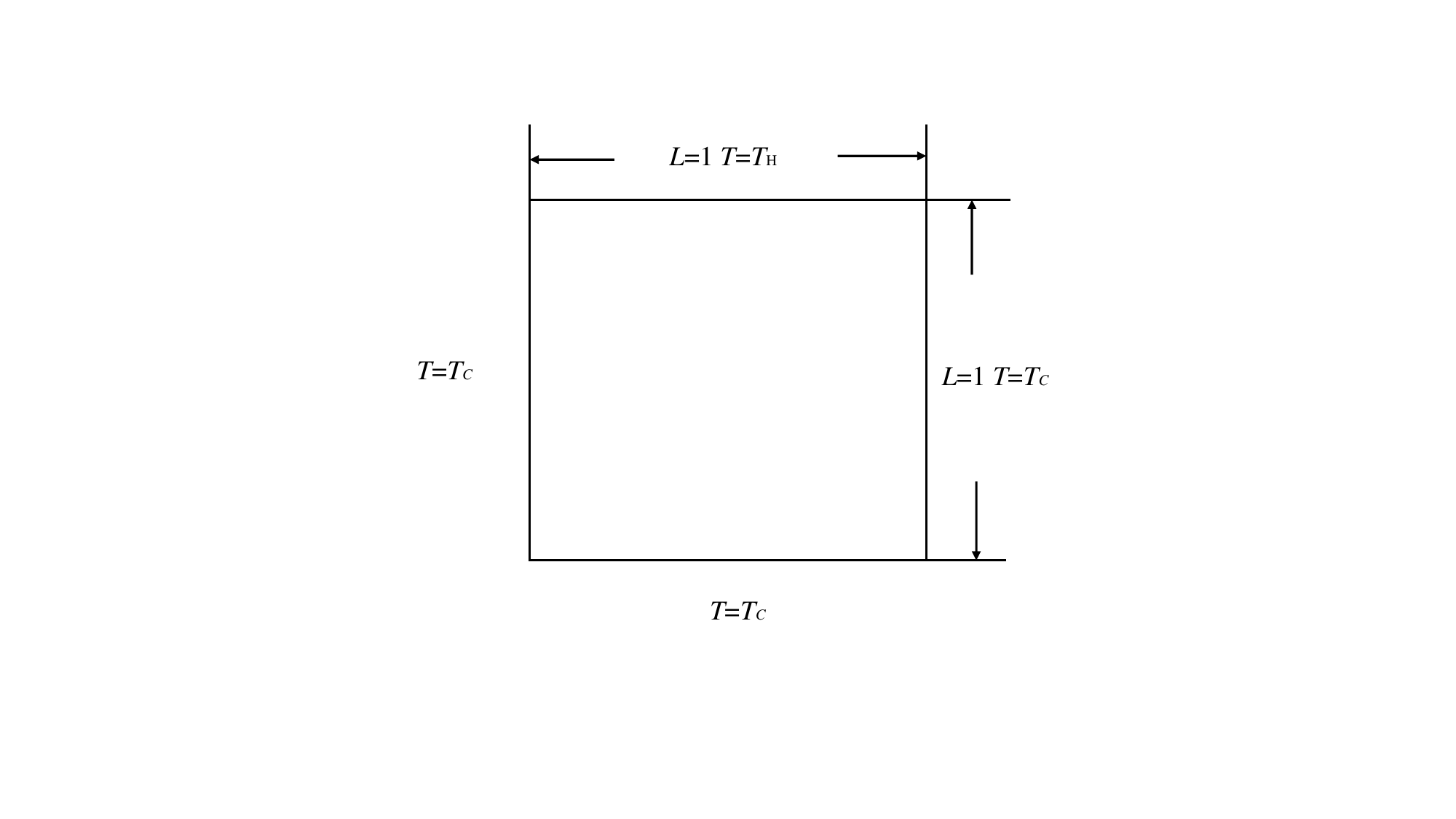}
	\caption{\label{2d-square}
		Computational domain and boundary condition of heat transfer in the 2D square domain. }
\end{figure}

In this case, we employ a two-dimensional uniform grid, discretized into 40 equally spaced points in each direction, resulting in a total of 1,600 grid points.
In this case, $N_{ref}=200$ in each cell for Kn = 10.0, 1.0, and $N_{ref}=50$ for Kn = 0.1, Kn = 0.01, balancing computational efficiency and statistical noise.
Furthermore, an additional 1000 steps were incorporated for statistical averaging in the two-dimensional simulation to reduce statistical noise.
The computation results for different Knudsen numbers are shown in Fig.\ref{2d-square-result}. The black solid lines in the figure represent the contour lines obtained using the UGKWP method, while the white dashed lines represent those computed with the DUGKS method. As can be seen, the UGKWP method exhibits very good agreement with the reference method in both the diffusive region and the ballistic region.

\begin{figure}[htb]	\label{2d-square-result}
	\centering	
    \includegraphics[height=0.40\textwidth]{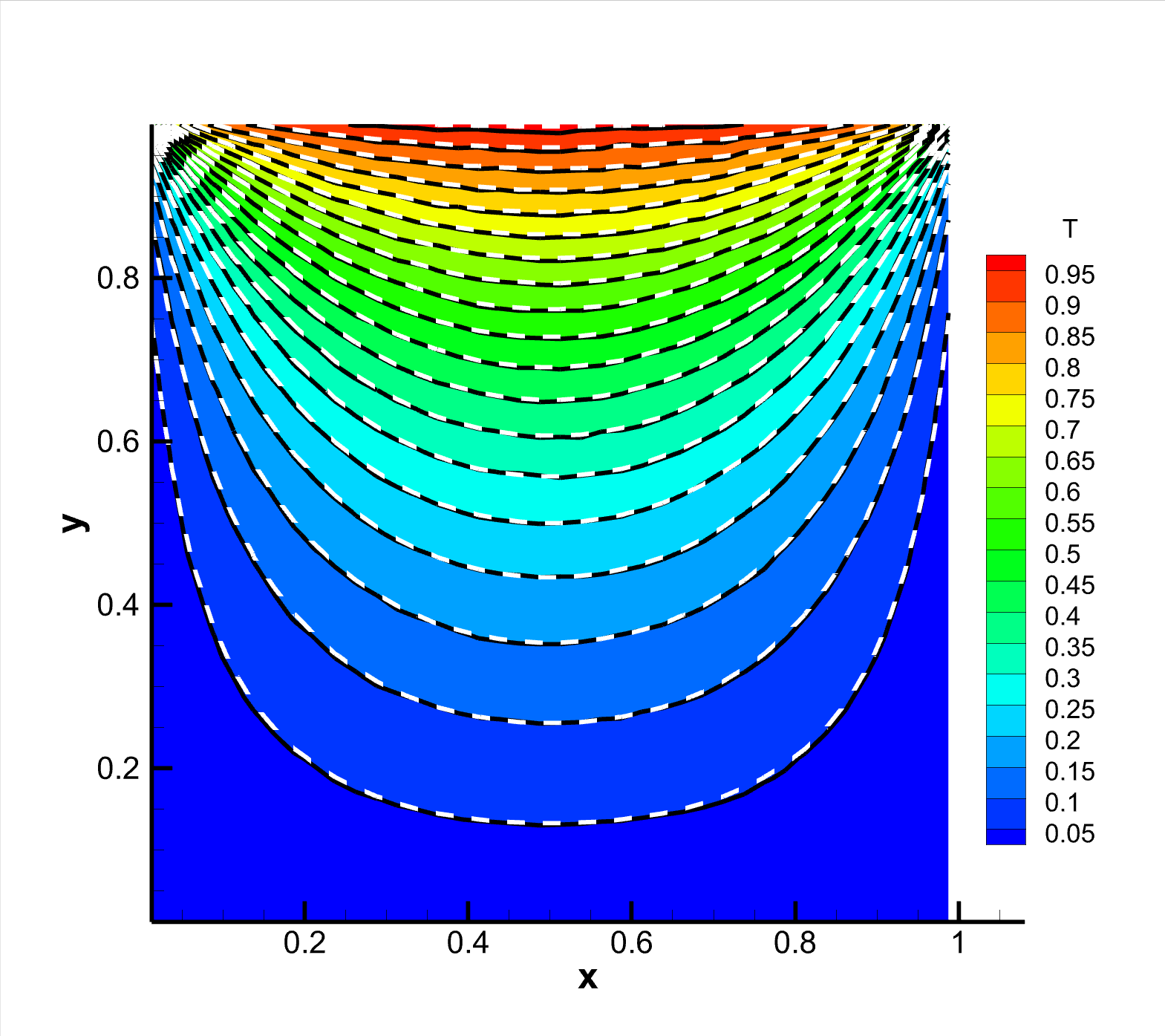}
	\includegraphics[height=0.40\textwidth]{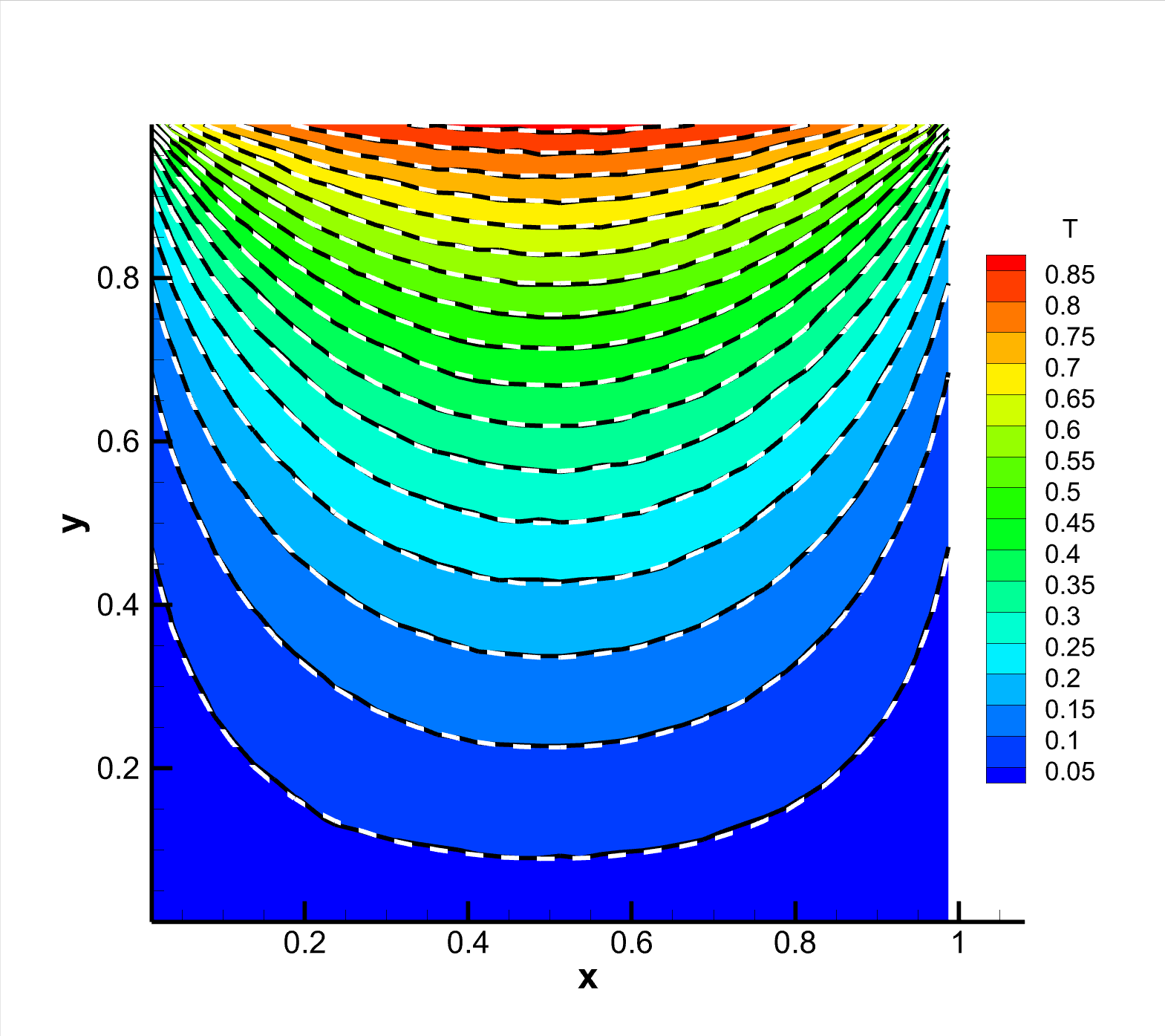}
    \includegraphics[height=0.40\textwidth]{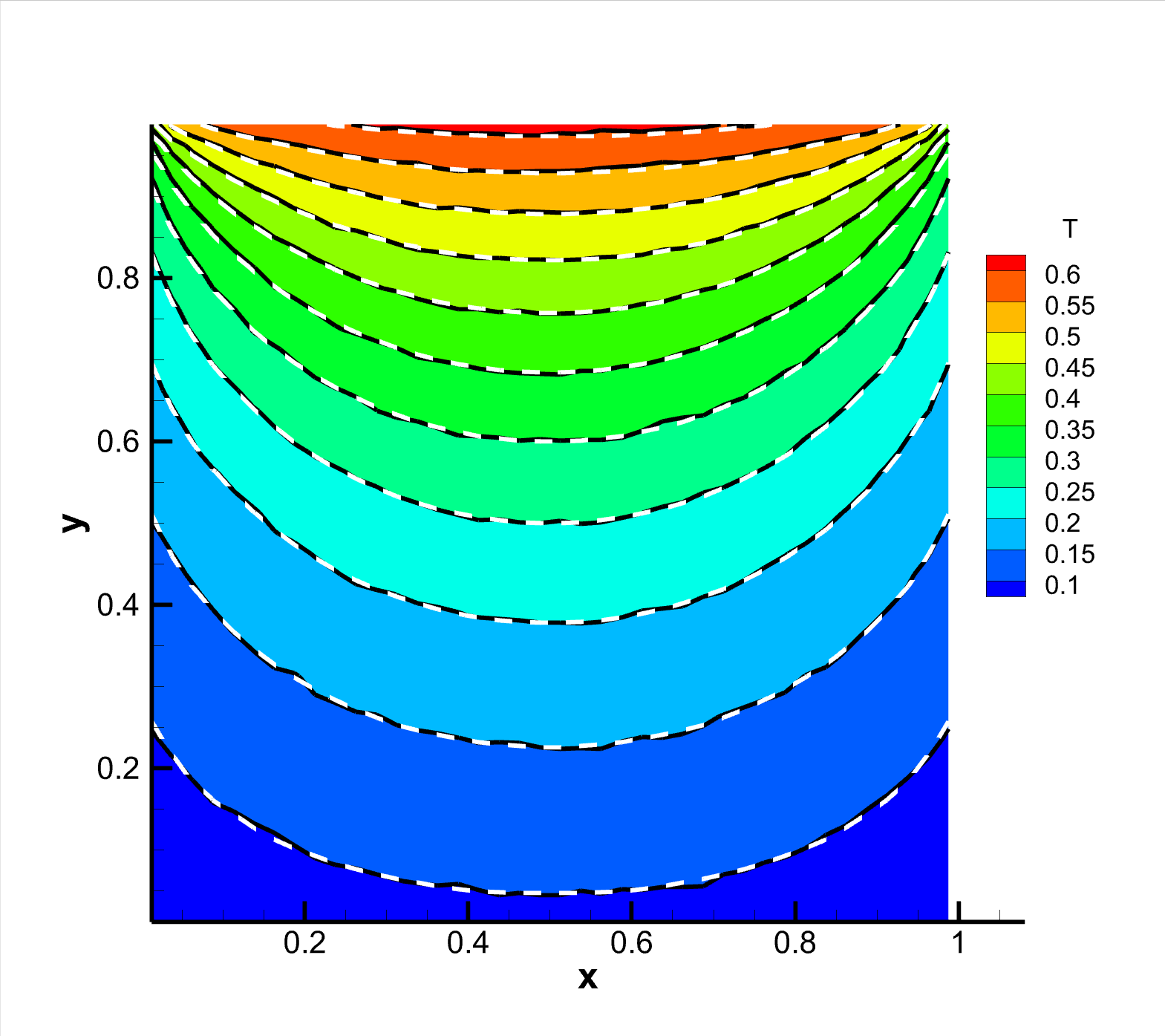}
    \includegraphics[height=0.40\textwidth]{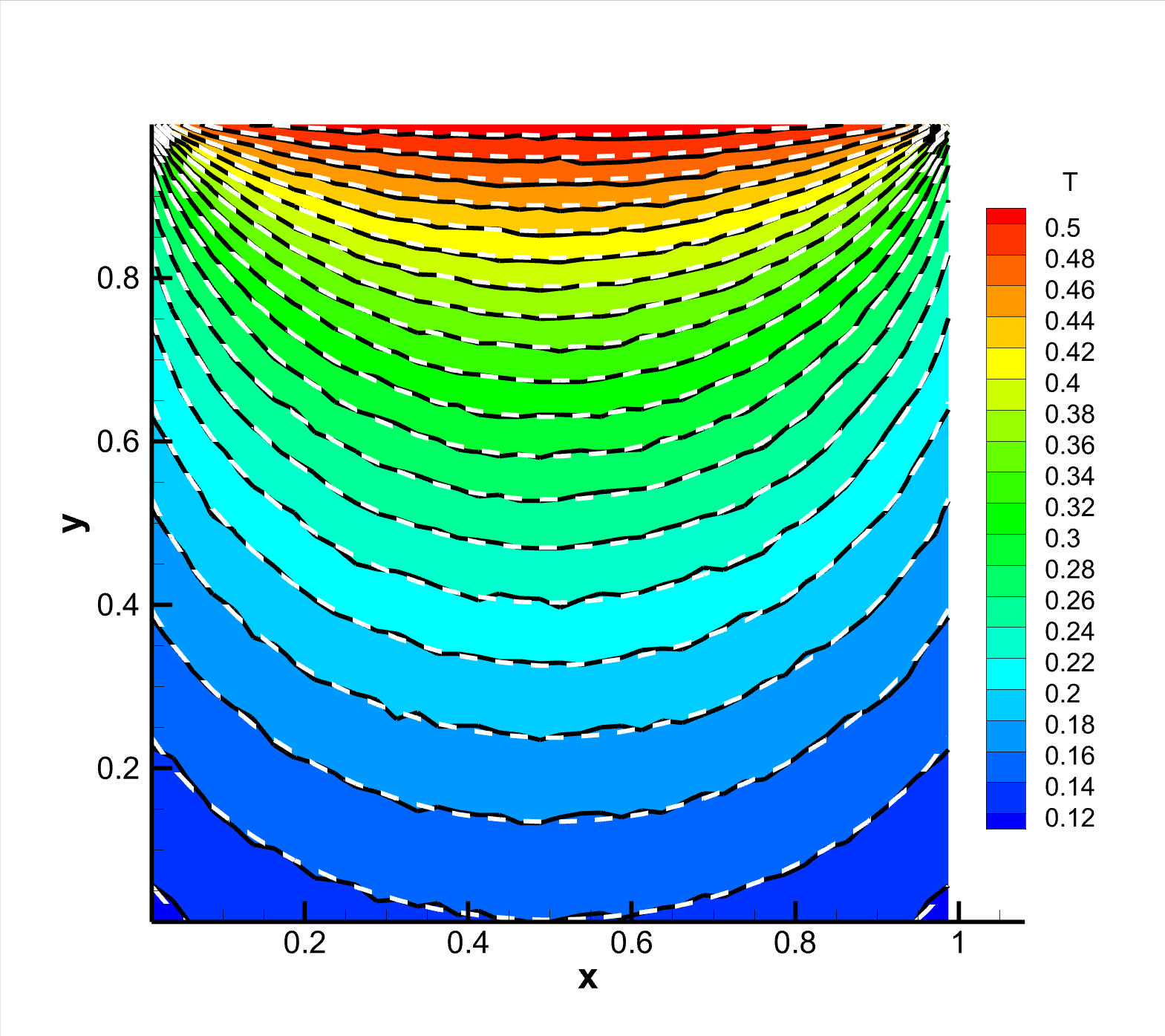}
	\caption{\label{2d-square-result}
		Comparison of 2D square heat transfer results, respectively corresponding: Kn = 0.01, Kn = 0.1, Kn = 1.0, Kn = 10.0. }
\end{figure}

\subsection{Heat transfer in the 2D rectangle domain}
This section investigates the heat transfer in a rectangular cavity with non-uniform temperature boundary conditions in the ballistic region. Computational domain and boundary conditions are shown in Fig.\ref{2d-rec}
\begin{figure}[htb]	\label{2d-rec}
	\centering	
	\includegraphics[height=0.50\textwidth]{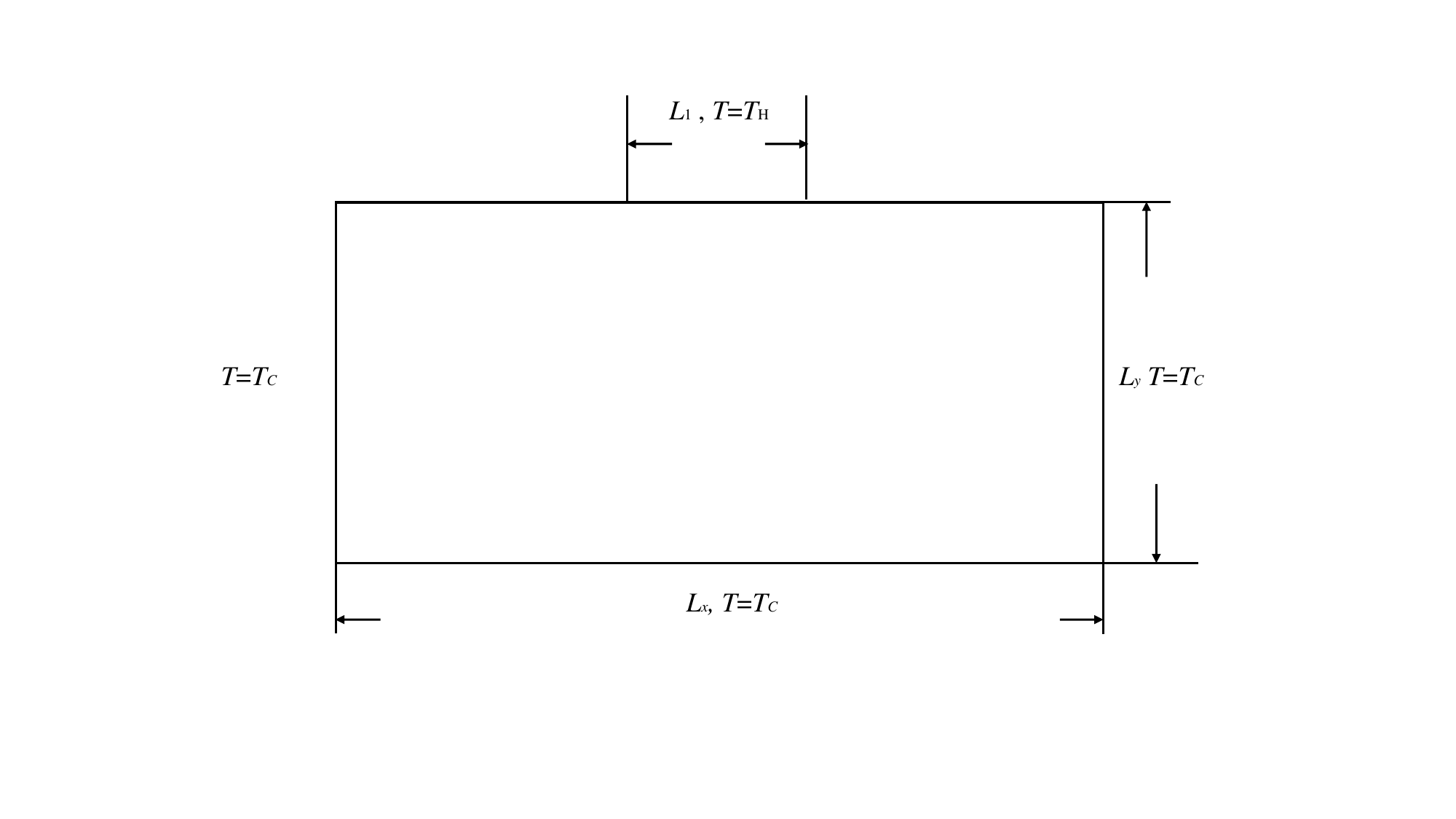}
	\caption{\label{2d-rec}
		Computational domain and boundary condition of Heat transfer in the 2D rectangle domain. }
\end{figure}
where $L_x=2L_y=5L_1$. The heat source is positioned at the center of the upper boundary of the entire computational domain. It has a length of $L_1 =1$ and a temperature of $T_H$. Other boundaries' temperature is $T_C$.

In this computational domain, the x-direction is discretized into 100 grid cells. In comparison, the y-direction is discretized into 50 grid cells which means the whole domain is discretized into 5000 uniform grid cells.
Meanwhile, the reference sampling number for each particle in each cell is 200 to balance efficiency and statistical noise.
Moreover, like heat transfer in the 2D square domain, in this case, a 1500-statistic average step is also adopted to reduce statistic noise.

The computation results for $Kn =10.0$ and $Kn = 1.0$ are shown in Fig.\ref{2d-square-result}. The black solid lines in the figure represent the contour lines obtained using the UGKWP method, while the white dashed lines represent those computed with the DUGKS method. As can be seen, the UGKWP method exhibits very good agreement with the reference method.

\begin{figure}[htb]	\label{2d-square-result}
	\centering	
    \includegraphics[height=0.40\textwidth]{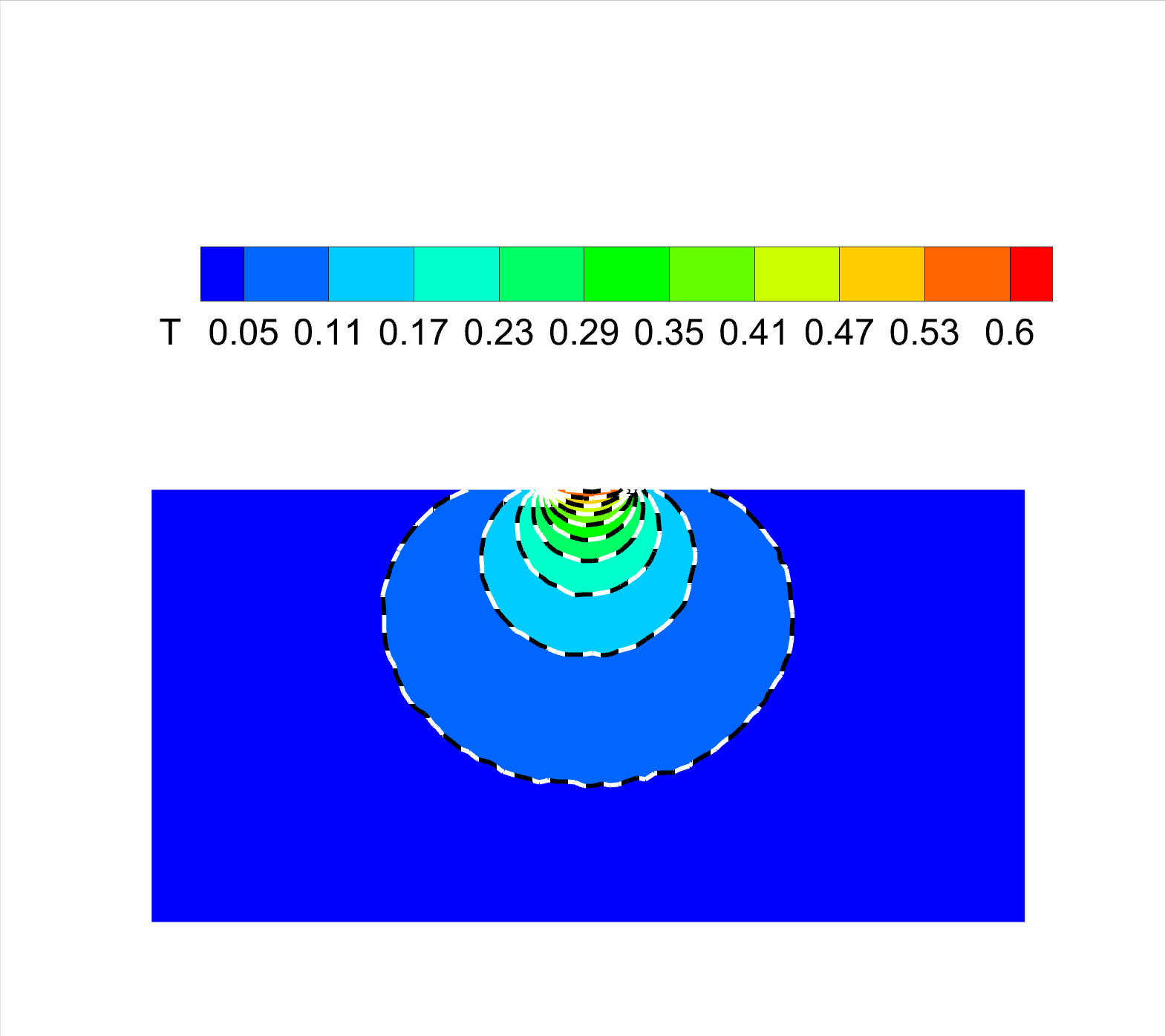}
	\includegraphics[height=0.40\textwidth]{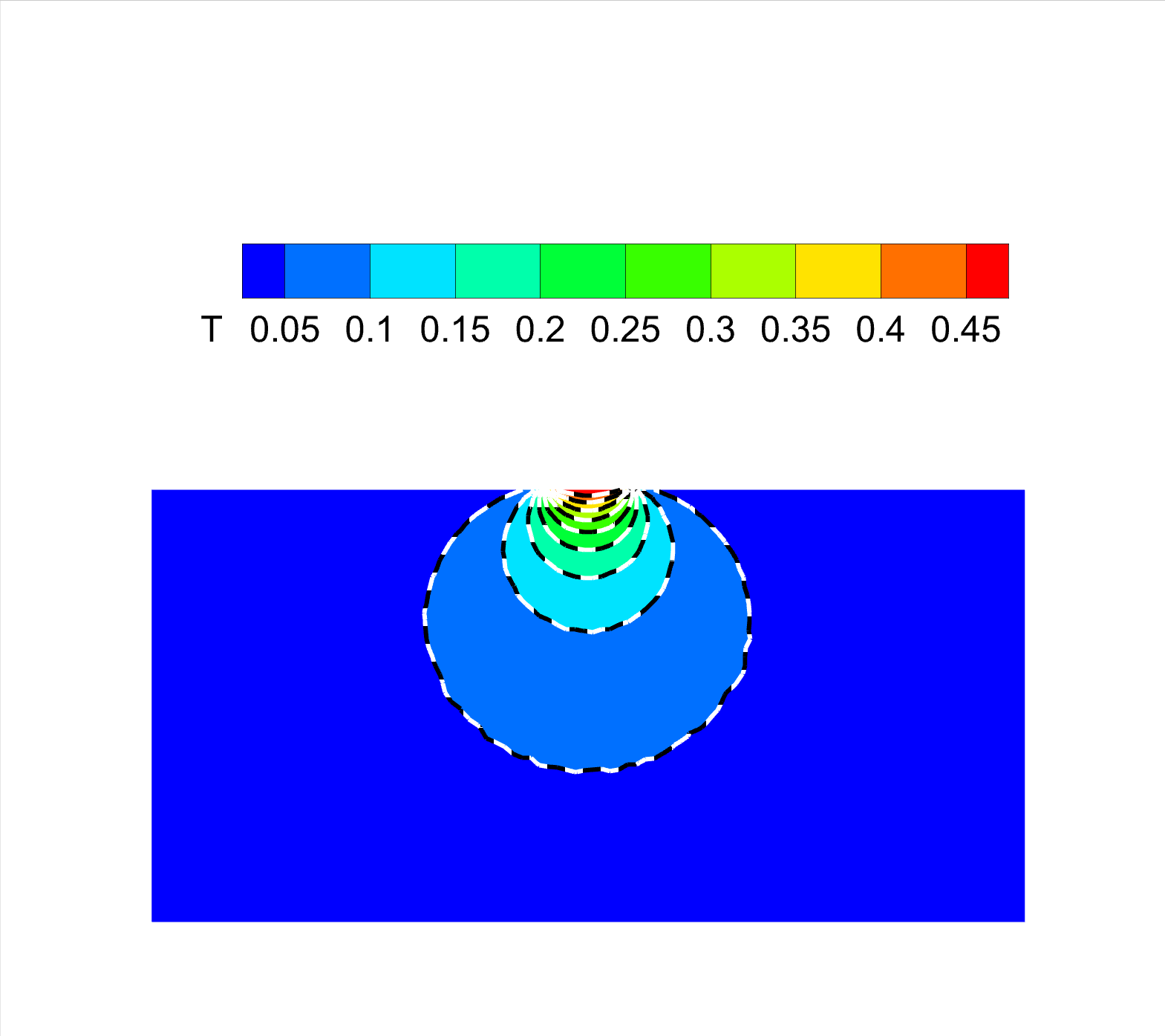}
	\caption{\label{2d-square-result}
		Comparison results of 2D rectangle heat transfer. Left: Kn = 1. Right: Kn = 10 }
\end{figure}

\subsection{2D multiscale heat transfer}
To further validate the UGKWP method's capability in capturing multi-scale non-equilibrium phonon transport in multi-dimensional problems, this section presents a domain-discontinuity problem. In region A, the Knudsen number is set to 10, corresponding to the ballistic regime, while in region B, it is 0.1, which is close to the diffusive regime as shown in Fig.\ref{2d-multi-domain}. The interface between the two regions exhibits a 100-fold difference in the Knudsen number, making this an excellent test case for multi-scale methods.

\begin{figure}[htb]	\label{2d-multi-domain}
	\centering	
    \includegraphics[height=0.50\textwidth]{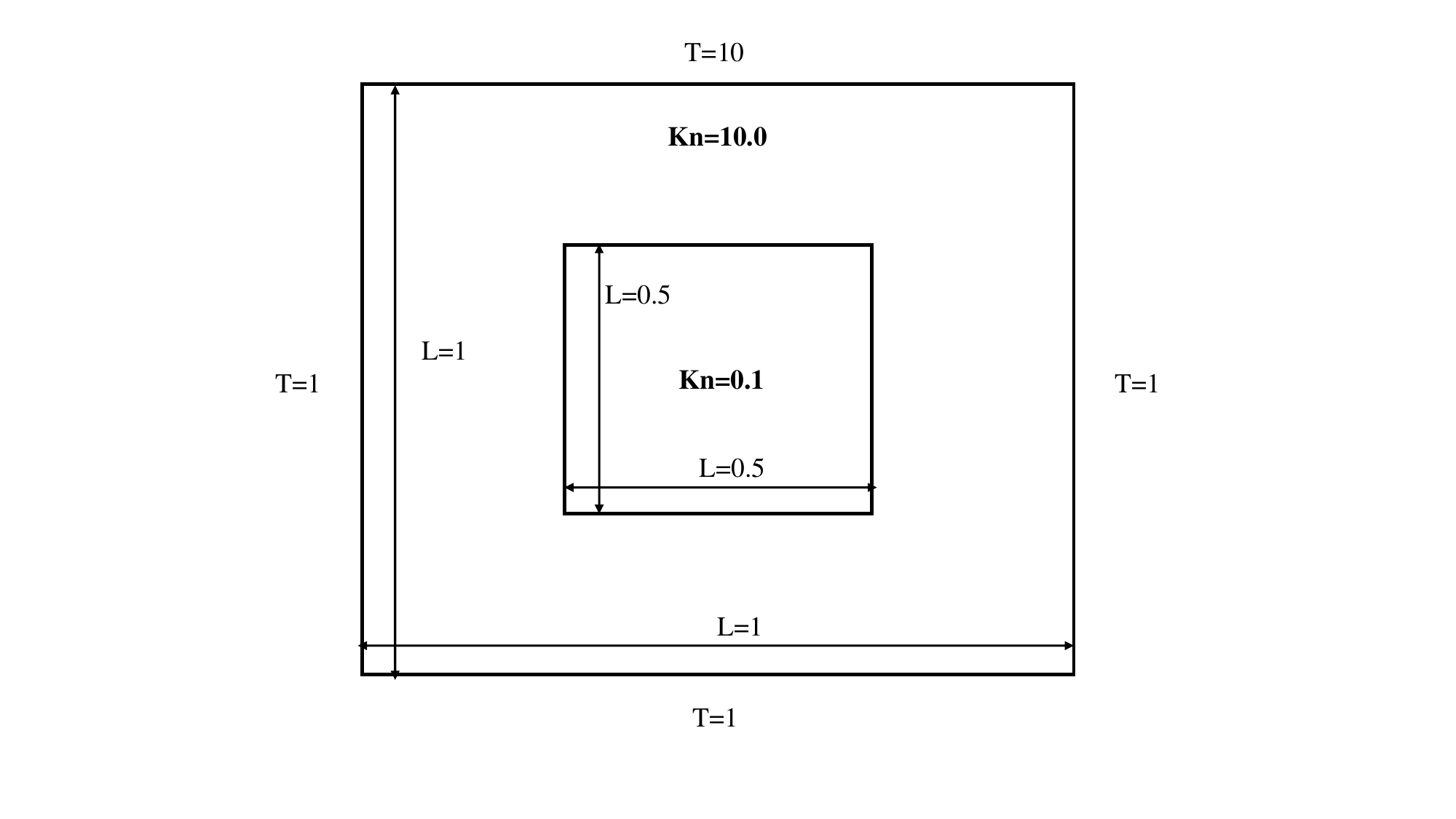}
	\caption{\label{2d-multi-domain}
		Computational domain and boundary condition for 2D multi-scale heat transfer. }
\end{figure}

In this computational domain, the x-direction is discretized into 80 grid cells. In comparison, the y-direction is discretized into 80 grid cells, which means the whole domain is discretized into 1600 uniform grid cells.
Meanwhile, the reference sampling number for each particle in each cell is 200 to balance efficiency and statistical noise.
Moreover, like heat transfer in the 2D square domain, in this case, a 1500-statistic average step is also adopted to reduce statistic noise.

The computational results, compared with those obtained using the UGKS method, are shown in the figure. In Fig.\ref{2d-multi-result}, the colored contour plot represents the UGKWP results, while the black contour lines represent the UGKS results. The comparison demonstrates that the UGKWP method agrees well with the reference method, exhibiting outstanding capability in capturing multi-scale non-equilibrium phenomena.

\begin{figure}[htb]	\label{2d-multi-result}
	\centering	
    \includegraphics[height=0.40\textwidth]{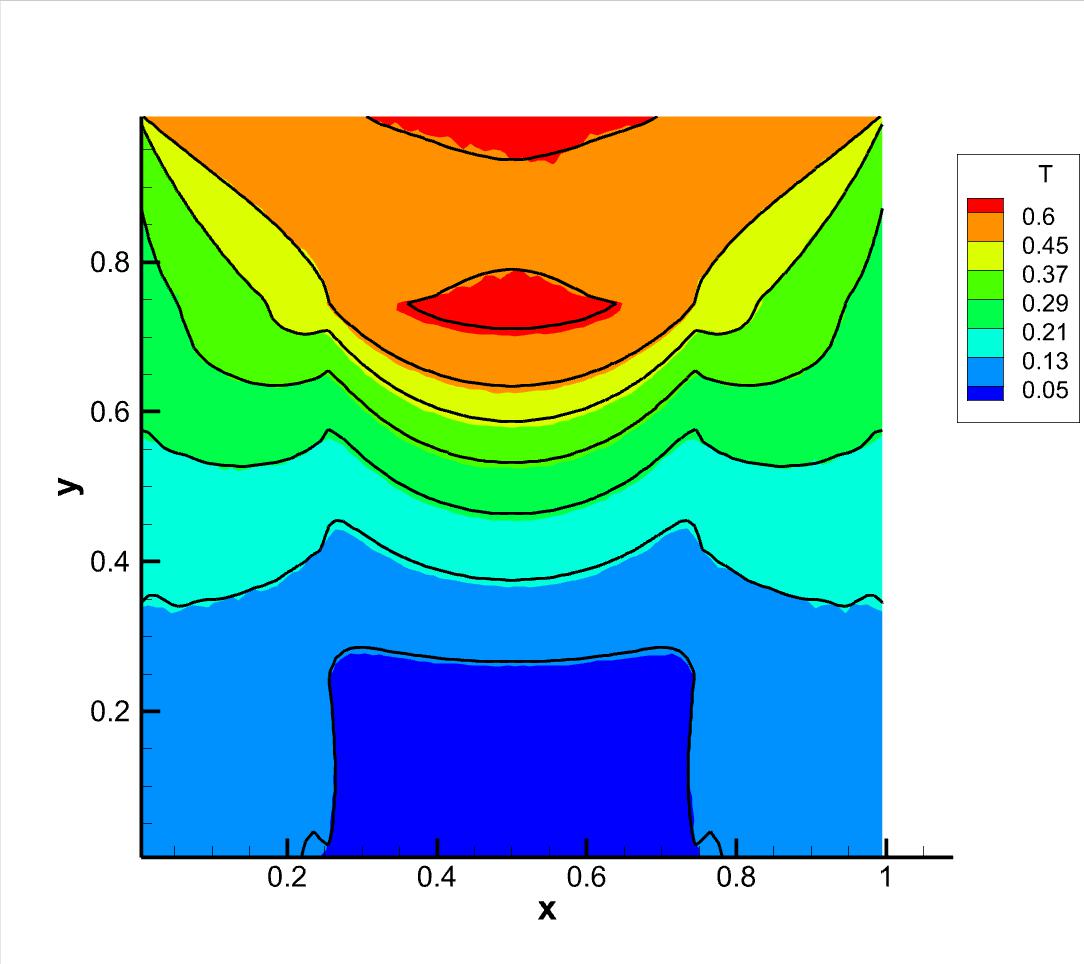}
    \includegraphics[height=0.40\textwidth]{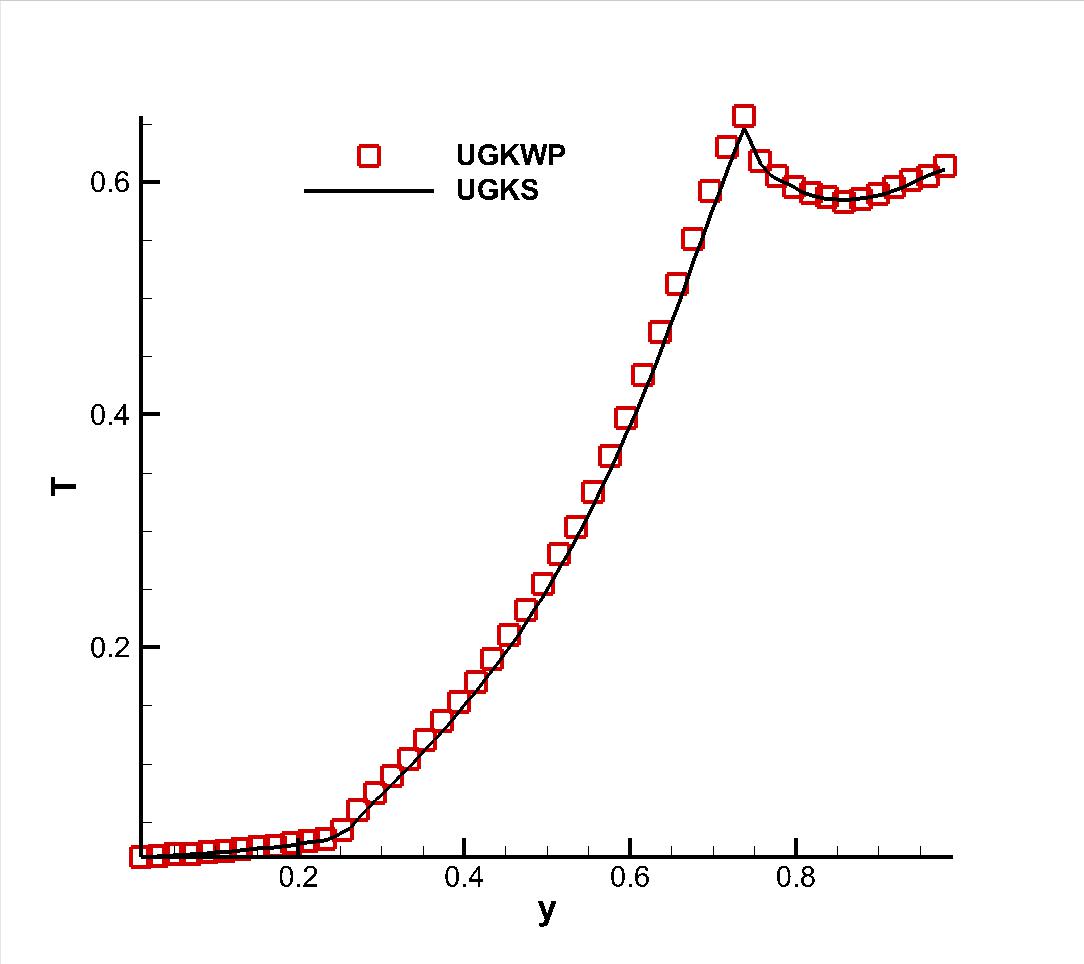}
	\caption{\label{2d-multi-result}
		Comparison results of 2D multi-scale heat transfer: (left) the temperature contour, (right) temperature distribution on the middle y-axis. }
\end{figure}

\subsection{Comparison of Computational Overhead: UGKS versus UGKWP}
In this section, we compare the computational cost of the UGKWP method with that of the UGKS method, demonstrating that the developed phonon transport UGKWP method offers a significant computational efficiency advantage over deterministic methods. We use an Intel i7 1260-p CPU to run the UGKS and the UGKWP method.

As reference particle number of the UGKWP method in heat transfer in the 2D square domain case, $N_{ref}=200$ in each cell for Kn = 10.0, 1.0, and $N_{ref}=50$ for Kn = 0.1, Kn = 0.01.
Meanwhile, the discrete velocity space number for UGKS is 64 x 64 for Kn = 10.0, 1.0, and 32 x 32 for Kn = 0.1, Kn = 0.01. The CPU time costs for both methods of 100 iterations are listed below:

\begin{table}[htb]
	\small
	\begin{center}
		\def\temptablewidth{1.0\textwidth}
		{\rule{\temptablewidth}{1pt}}
		\begin{tabular*}{\temptablewidth}{@{\extracolsep{\fill}}c|c|c|c|c}
			Scheme & Kn = 10.0 & Kn =1.0  & Kn = 0.1  & Kn =0.01 \\
			\hline
			UGKWP 	&41s & 45s  & 12s & 9s \\ 	UGKS 	&839s & 864s  & 212s & 208s \\
		\end{tabular*}
		{\rule{\temptablewidth}{0.1pt}}
	\end{center}
	\vspace{-4mm} \caption{\label{viscous subsonic sphere} CPU time cost comparisons among UGKWP and UGKS for the heat transfer in the 2D square domain.}
\end{table}

\section{Conclusion}

A UGKWP method is developed in this paper for solving the transient phonon BTE, which can seamlessly adapt to heat conduction problems across a broad range of Knudsen numbers.
In the diffusive limit, the method naturally recovers Fourier's law of heat conduction as the particle contribution becomes negligible. Conversely, in the ballistic limit, the non-equilibrium flux is entirely characterized by free-streaming particles.
A series of numerical tests spanning the diffusive to ballistic regimes demonstrates that the UGKWP method produces results that agree well with benchmark solutions and possesses remarkable multi-scale adaptability and versatility, effectively bridging the gap between diffusive and ballistic transport.

One limitation of the current method is that it exhibits higher levels of statistical noise in the ballistic regime due to its reliance on statistical particles to capture non-equilibrium effects.
However, this issue can be effectively mitigated through statistical averaging, as demonstrated in the two-dimensional examples, where the results showed nearly no statistical noise.
In future work, the method can also be extended to unstructured meshes and anisotropic phonon dispersions obtained by $ab~initio$ \cite{zhang2023acceleration}.

\section{Acknowledgements}

The current research is supported by National Key R\&D Program of China (Grant Nos. 2022YFA1004500), National Science Foundation of China (12172316, 92371107), and Hong Kong research grant council (16301222, 16208324).

\bibliographystyle{unsrt}
\bibliography{jixingbib}

\end{document}